%% file: state-of-the-art.tex
\documentclass[11pt]{article}

\usepackage{amsmath}
\usepackage{bm}
\usepackage{bookman}
\usepackage[a4paper,width=16cm,height=24.7cm]{geometry}
\usepackage{graphicx}
\PassOptionsToPackage{hyphens}{url}
\usepackage[backref=false,colorlinks=true,final,linkcolor=magenta,urlcolor=blue,citecolor=black]{hyperref}
\usepackage[utf8]{inputenc}
\usepackage{multicol}
\usepackage{paralist}
\usepackage[norefs,nocites]{refcheck}	
\usepackage{subcaption}
\usepackage{titlesec}
\usepackage{xcolor}
\usepackage{authblk}

\titleclass{\subsubsubsection}{straight}[\subsection]

\newcounter{subsubsubsection}[subsubsection]
\renewcommand
  \thesubsubsubsection{\thesubsubsection.\arabic{subsubsubsection}}

\titleformat{\subsubsubsection}
  {\normalfont\normalsize\bfseries}{\thesubsubsubsection}{1em}{}
\titlespacing*{\subsubsubsection}
{0pt}{3.25ex plus 1ex minus .2ex}{1.5ex plus .2ex}

\makeatletter
\renewcommand\paragraph{\@startsection{paragraph}{5}{\z@}%
  {3.25ex \@plus1ex \@minus.2ex}%
  {-1em}%
  {\normalfont\normalsize\bfseries}}
 
\def\toclevel@subsubsubsection{4}
\def\toclevel@paragraph{5}
\def\l@subsubsubsection{\@dottedtocline{4}{7em}{4em}}
\def\l@paragraph{\@dottedtocline{5}{10em}{5em}}
\makeatother

\title{\textbf{A Review of Formal Methods\\applied to Machine Learning}}
\author[1]{Caterina Urban}
\author[2,3]{Antoine Miné}
\affil[1]{Inria \& École Normale Supérieure $\mid$ Université PSL, F-75005 Paris, France 
\texttt{caterina.urban@inria.fr}}
\affil[2]{Sorbonne Université, CNRS, LIP6, F-75005 Paris, France \texttt{antoine.mine@lip6.fr}}
\affil[3]{Institut Universitaire de France, F-75005, Paris, France}

\begin{document}

\maketitle

\input{abstract}

\section{Introduction}

Recent advances in artificial intelligence and machine learning and the
availability of vast amounts of data allow us to develop computer software
that efficiently and autonomously perform complex tasks that are difficult
or even impossible to design using traditional explicit programming (e.g.,
image classification, speech recognition, etc.).
This makes machine-learned software particularly attractive even in
\emph{safety-critical applications}, as it enables performing a whole world
of new functions that could not be envisioned before, e.g., autonomous
driving in the automotive industry, or image-based operations (taxiing,
takeoff, landing) and aircraft voice control in the avionics industry.
Another attractive aspect of machine-learned software is its ability to
efficiently approximate or simulate complex processes and systems and
automate decision-making, e.g., diagnosis and drug discovery processes in
healthcare, or aircraft collision avoidance systems in avionics
\cite{ACASXu}.

Safety-critical applications require an extremely high level of insurance
that the software systems behave correctly and reliably.
Today, \emph{formal methods} are an integral part of the development process
of traditional (non machine-learned) critical software system, to provide
strong, mathematically-grounded guarantees on the software behavior.
For instance, they are used at an industrial level in avionics
\cite{Airbus}, where the development processes of aircraft software systems
have very stringent assurance and verification requirements mandated by
international standards (i.e., DO-178C).
This success is due to the recognition of formal methods by certification
authorities and the availability of effective and efficient verification
tools.
Among formal verification techniques, \emph{static analyzers} are
particularly successful as they are fully automated, efficient, and correct
by construction. A notable example is the Astrée static analyzer
\cite{Astree}, used to ensure absence of run-time errors in critical
avionics C code.

In contrast, research in formal methods for the development and assurance of
machine learning systems remains today extremely limited. It is however
rapidly gaining interest and popularity, mostly driven by the growing needs
of the automotive and avionics industries.
The purpose of this document is to keep up with these developments and gain
a better understanding of the research ecosystem that is forming around the
verification of machine learning software, and discuss further
research directions.
Specifically, in the following, we give an introduction to formal methods (Section~\ref{sec:fm})
with a particular focus on static analysis by \emph{abstract interpretation}
\cite{CC77}, as it offers a unifying theory for reasoning about disparate formal verification approaches (Section~\ref{sec:fm:ai}). We then thoroughly overview the current
state of the art in formal methods for machine learning
(Section~\ref{sec:soa}). We provide descriptions of the different underlying
techniques and discuss and compare the scope of their application and their
advantages and disadvantages.
Finally, we discuss perspectives and expectations for possible worthwhile future
research directions (Section~\ref{sec:future}).

\section{Formal Methods}\label{sec:fm}

\input{fm.tex}

\section{State of the Art}\label{sec:soa}

\vspace{-1em}
\begin{figure}[ht]
	\centering
	\includegraphics[width=\textwidth]{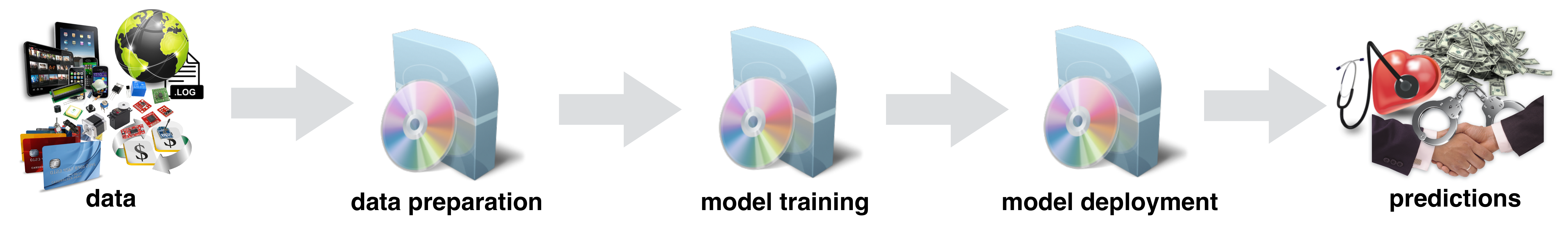}
	\caption{Machine Learning Pipeline.}\label{fig:pipeline}
\end{figure}

\noindent
The main differences between machine learning software and
traditional software arise from the software development process. The
\emph{machine learning pipeline} (cf. Figure~\ref{fig:pipeline})
begins with the collection of data and its preparation for training,
followed by a machine learning model training phase, and ends with the
deployment of the trained model and its use to make predictions and
automate tasks.
The training phase is highly non-deterministic, which is in contrast with
predictability and traceability requirements of traditional safety-critical
software development processes. Furthermore, the trained models only give
probabilistic guarantees on the prediction results, bound to the data used
for training, which are not sufficient for guaranteeing an acceptable
failure rate under any circumstances, as required for traditional critical
software. Thus, formal verification methods need to adapt to these
differences in order to ensure the safe development and deployment of
machine learning software in safety-critical applications.
Most formal verification methods proposed so far in the literature apply to
\emph{trained machine learning models} (Section~\ref{sec:models}) while
only few are dedicated to the \emph{data preparation} and \emph{model
training} phases (Section~\ref{sec:other}).

\subsection{Formal Methods for Trained Machine Learning Models}
\label{sec:models}

So far, the vast majority of the research in verification of machine
learning models has focused on \emph{neural networks} \cite{DL}
(Section~\ref{sec:dl}), driven by their success and the interesting
research challenges they bring for the formal methods community.
Traditional machine learning models such as \emph{decision trees} \cite{DT}
have been initially left aside as they are much easier to verify than
neural networks \cite{Bastani2} and thus not considered interesting enough.
This is not necessarily the case for decision tree ensemble models, such as
\emph{random forests} \cite{RF} and \emph{gradient boosted decision trees}
\cite{GBDT}. Indeed, a small but growing number of recent formal
verification methods target these models as well as \emph{support vector
machines} \cite{SVM} (Section~\ref{sec:ml}).

\subsubsection{Formal Methods for Neural Networks}
\label{sec:dl}

\begin{figure}[ht]
	\centering
	\includegraphics[width=0.75\textwidth]{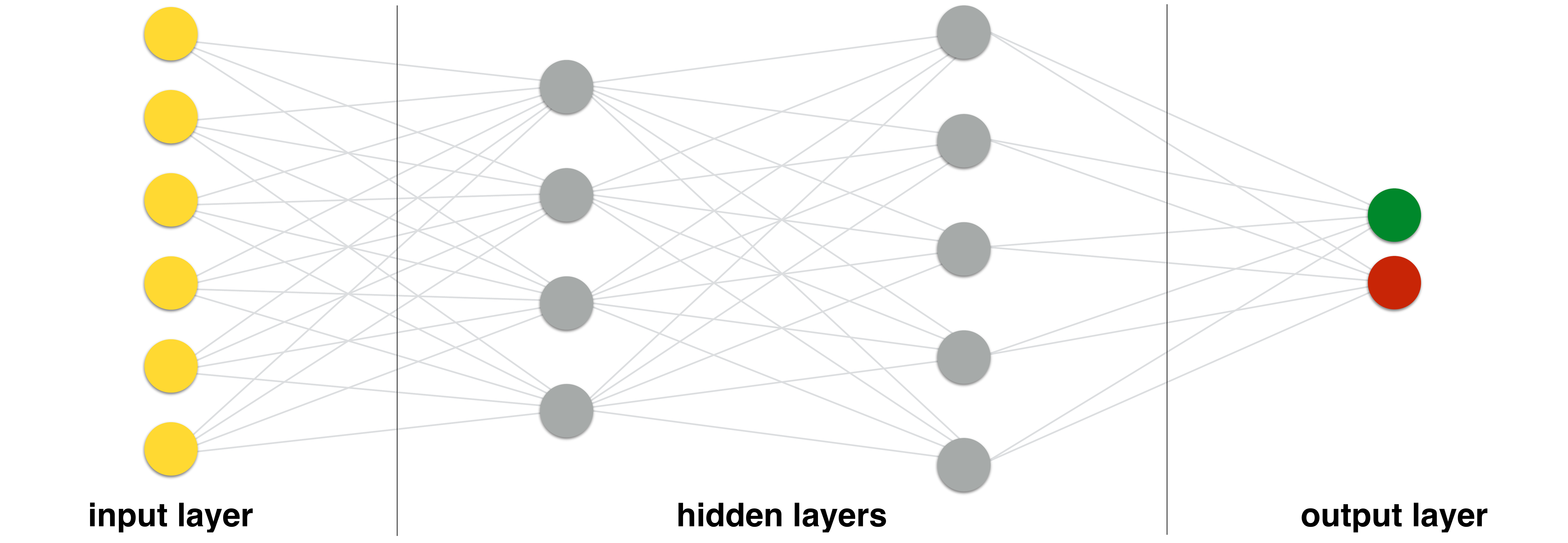}
	\caption{Feed-Forward Fully-Connected Neural Network.}\label{fig:nn}
\end{figure}

\noindent
Neural networks \cite{DL} are collections of connected nodes
called \emph{neurons}. Each neuron acts as a computational unit: it
receives inputs or signals from other neurons, processes them, and
transmits the resulting signal to other connected neurons. Connections
between neurons are assigned a weight which increases or decreases the
strength of the signal. Specifically, neurons apply a (non-linear)
\emph{activation function} to the (weighted) sum of their inputs to produce
their output signal. 
Nowadays, the most commonly used activation function is the \emph{Rectified
Linear Unit (ReLU)} \cite{ReLU}, i.e., $ReLU(x) = max\{0, x\}$. Other
popular activations include the \emph{sigmoid or logistic function}, i.e.,
$\sigma(x) = \frac{1}{1 + e^{-x}}$, and the \emph{hyperbolic tangent
function (tanh)}, i.e., $tanh(x) = \frac{e^x - e^{-x}}{e^x + e^{-x}}$.

Neurons are typically aggregated into \emph{layers} such that neurons in
one layer only connect to neurons in the (usually immediately) following
layer. Multiple connection patterns are possible between layers, which
define different layer types. In a \emph{fully-connected layer} each neuron
receives inputs from every neuron in the previous layer, while in a
\emph{pooling layer} or in a \emph{convolutional layer} each neuron only
receives inputs from a subset of neurons in the previous layer. Neural
networks that only consist of fully-connected and pooling layers are known
as \emph{feed-forward neural networks} (cf. Figure~\ref{fig:nn}), while
neural networks with at least one convolutional layer are called
\emph{convolutional neural networks}. These neural networks form directed
acyclic graphs in which signals travel from the input layer, through each
of the hidden layers, to the output layer. Instead, \emph{residual neural
networks} use shortcuts to jump over some layers, and \emph{recurrent
neural networks} allow neurons to connect to neurons in previous layers
(thus forming cycles in which signals traverse layers multiple times).

Kurd and Kelly \cite{Kurd} were the among the first to propose a
characterizations of verification goals for neural networks used in
safety-critical applications. Most existing formal methods for neural
networks aim at verifying what Kurd and Kelly identify as goals G4 and G5,
which respectively concern ensuring robustness to disturbances to inputs
and ensuring that outputs of neural networks are not hazardous. In the
following, in addition to the targeted verification goal, we also
categorize methods based on the supported neural networks (i.e.,
architecture, activation functions, size, etc.) and the underlying
technique used. We broadly distinguish between \emph{complete verification methods}
(Section~\ref{sec:complete}) and \emph{incomplete verification methods}
(Section~\ref{sec:incomplete}).

\subsubsubsection{Complete Formal Methods}
\label{sec:complete}

Complete formal verification methods generally \emph{do not scale} to large
neural network architectures (they generally require several hours for
neural networks with hundreds or thousands of neurons) but are both
\emph{sound} and \emph{complete}: they can precisely report whether or
not a given property holds on a neural network, generally providing
a counter-example in the later case.
Note, however, that soundness is not typically guaranteed with respect to
floating-point arithmetic, but only with respect to computations on reals
that ignore rounding errors and may thus differ from the actual computations
\cite{Neumaier,Jiangchao}.
Complete methods are often also limited to certain neural network
architectures and activations (e.g., neural networks with piecewise linear
layers and ReLU activations).
\looseness=-1
We distinguish below between methods based on \emph{satisfiability modulo
theory (SMT) solving} \cite[etc.]{Pulina10,Planet,Reluplex,Marabou,DLV},
methods based on \emph{mixed integer linear programming (MILP)}
\cite[etc.]{Cheng1,Fischetti,BaB,Sherlock,MIPVerify}, and other complete
methods based on global optimization \cite{DeepGO} or combinations of exact
and approximate analyses
\cite[etc.]{ReluVal,Neurify,Polyhedron,StarSets,RefineZono}.

\paragraph{SMT-based Formal Methods.}

SMT-based formal methods reduce the safety verification problem to a
constraint satisfiability problem. Specifically, they encode a given neural
network and (the negation of a) desired safety property as a set of
constraints. If the constraints are satisfiable, the corresponding solution
represents a counterexample to the safety property. Otherwise, the
constraints are unsatisfiable, no counterexample can exist and thus the
safety property holds.
For instance, consider a $n$-layer single-output feed-forward neural
network with ReLU activations after each hidden layer, and a safety
property specifying a bounded input domain and requiring the output of the
neural network to be a positive value. The verification of the safety
property can be reduced to the following SMT problem \cite{BaB}:
\begin{subequations}
\begin{align}
	\label{eq:smt-input}
	&\mathbf{l} \leq \mathbf{x}_0 \leq \mathbf{u} \\
	\label{eq:smt-layer}
	&\mathbf{\hat{x}}_{i+1} 
	= \mathbf{W}_{i+1} \mathbf{x}_i + \mathbf{b}_{i+1} 
	& \forall i \in \{0, \dots, n-1\}\\
	\label{eq:smt-relu}
	&\mathbf{x}_i = max\{0, \mathbf{\hat{x}}_i\} 
	& \forall i \in \{1, \dots, n-1\} \\
	\label{eq:smt-output}
	&\mathbf{x}_n \leq 0
\end{align}
\end{subequations}
where Equation (\ref{eq:smt-input}) and Equation (\ref{eq:smt-output})
respectively represent the constraints on the inputs and (the negation of
those) on the output of the neural network, while Equation
(\ref{eq:smt-layer}) encodes the affine transformations performed by the
layers of the network, and Equation (\ref{eq:smt-relu}) encodes the ReLU
activation functions. Note that the presence of these non-linear ReLU
constraints makes the problem NP-complete \cite{Reluplex}. A value
assignment to the problem variables that satisfies all constraints
represents a valid counterexample. If the problem is unsatisfiable, the
safety property is verified.

\mbox{}

The first formal verification method for neural networks was presented by
\textbf{Pulina and Tacchella} \cite{Pulina10}.
They proposed an approach for checking that the output of a feed-forward
fully-connected neural network with sigmoid activations always ranges
within given safety bounds.
The neural network is encoded as a Boolean combination of linear
constraints using a Cartesian product of intervals. Specifically,
activation functions are replaced by piecewise linear approximations using
$p$ intervals (where $p$ is a parameter of the approach). The encoding of
the neural network together with the constraints encoding the desired
output safety property are then fed to a black-box SMT solver. When spurious
counterexamples are found, the encoding is refined to use $p / r$ intervals
(where $r$ is the refinement rate, another parameter of the approach).
The approach was implemented and evaluated on a small case study using a
neural network with one hidden layer containing three neurons. The authors
show that a $10x$ increase in precision corresponds to $100x$ increase in
the size of the encoding, thus highlighting the difficulty in scaling-up
this approach.

In follow-up work \cite{Pulina12}, Pulina and Tacchella perform a
comparison of different back-end SMT solvers for their approach on various
slightly larger neural networks with up to $20$ hidden neurons. They
conclude that the \textsc{Yices} SMT solver \cite{YICES} seems the best
candidate. Nonetheless, the scalability of the approach remains an issue.

A similar approach to that of Pulina and Tacchella is presented by
Scheibler et al. \cite{Scheibler}. It is based on bounded model checking
\cite{BMC} and applied to the cart-pole (or inverted pendulum) system. They
also encounter scalability issues and can only handle small unrolling
depths and very basic safety properties.

\mbox{}

\begin{figure}[ht]
	\centering
	\includegraphics[width=0.45\textwidth]{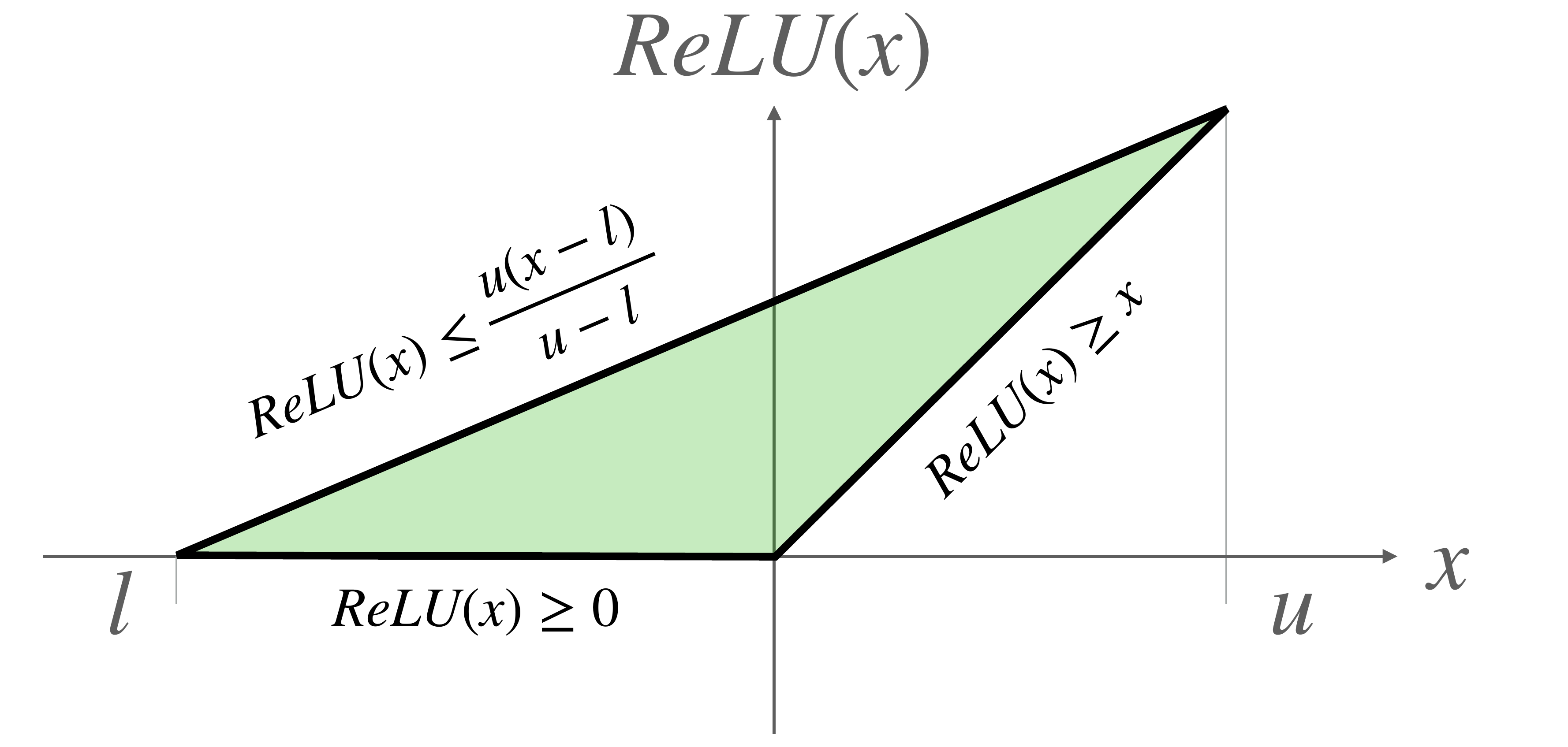}
	\caption{Convex Approximation of a ReLU Activation.}\label{fig:triangle}
\end{figure}

More recently, Ehlers \cite{Planet} proposed an approach for proving convex
input-output properties of neural networks with piecewise linear layers
(i.e., fully-connected, max-pooling, and convolutional layers) and ReLU
activations. The approach is implemented in a tool named \textbf{Planet}
(\url{https://github.com/progirep/planet}).
The underlying idea of the approach is to use approximations in order to
reduce the search space for the SMT solver.
In a first step, Planet uses interval arithmetic \cite{IntervalArithmetic}
to obtain lower and upper bound estimations for each neuron.
Afterwards, Planet encodes the behavior of the neural network as a
combination of linear constraints. ReLU activations are approximated using
the three linear constraints shown in Figure~\ref{fig:triangle}, where $l$
and $u$ are the lower and upper bounds for a neuron $x$ and $ReLU(x)$ is
the result of the activation function.
Finally, the approximation of the neural network is conjoined together with
the (negation of the) desired safety property to encode the verification
problem. The search for solutions is done by fixing the activation status
of the ReLU activations (i.e., adding the constraint $x > 0$ to make
$ReLU(x)$ always active, $ReLU(x) = x$, or adding the constraint $x \leq 0$
to make $ReLU(x)$ always inactive, $ReLU(x) = 0$) and backtracking when
needed (i.e., when constraints become unsatisfiable).
The approach is evaluated on a small vehicle collision avoidance use case
and on a neural network with over a thousand neurons trained for digit
recognition on the MNIST dataset \cite{MNIST}. The scalability
considerably improves compared to previous works but the approach still
timeouts (after four hours) for difficult verification properties (e.g.,
determining how much noise can be added to images to make the neural
network misclassify them).

\mbox{}

Katz et al. \cite{Reluplex} proposed a custom SMT solver named
\textbf{Reluplex} for verifying safety properties of feed-forward
fully-connected neural networks with ReLU activations.
The underlying approach is based on the simplex algorithm \cite{Dantzig},
extended to support ReLU constraints.
Given an SMT encoding of the verification problem (cf. Equations
(\ref{eq:smt-input})--(\ref{eq:smt-output})), Reluplex always maintains a
value assignment for all variables, even if some constraints of the
encoding are violated. At each step, the assignment is modified to fix some
violated constraints. To ensure convergence, ReLU constraints that get
fixed too often are split into two cases, each corresponding to one
activation status of the ReLU. In the worst case, the problem will be split
over all possible combinations of activation patterns into simple linear
problems. If no valid assignment that satisfies all constraints can be
found, the safety property is satisfied.
The approach is applied to the verification of the ACAS Xu neural networks
\cite{ACASXu}, developed as early prototype for the next-generation
airborne collision avoidance system for unmanned aircraft. The neural
networks take sensor data as input (i.e., speed and present course of the
current aircraft and of any nearby intruder) and issue appropriate
navigation advisories. They consist of six hidden layers with $300$ neurons
each. Reluplex requires several hours to perform the verification (over two
days in one case).

\textbf{Marabou} \cite{Marabou}
(\url{https://github.com/NeuralNetworkVerification/Marabou}) is the
successor of Reluplex.
It supports neural networks with piecewise linear layers (i.e.,
fully-connected, max- and average-pooling, and convolutional layers) and
piecewise linear activations.
Unlike Reluplex, Marabou offers a divide-and-conquer solving mode, which
iteratively splits the input space and thus naturally lends itself to
parallel execution.
Additionally, Marabou maintains symbolic lower and upper bounds for all
neurons expressed as linear combinations of the input neurons
\cite{ReluVal}. This allows Marabou to determine tighter bounds and further
pruning of the input space.
Finally, Marabou includes several implementation improvements over
Reluplex. For instance, it integrates a custom simplex solver instead of
relying on an external solver.
The experimental evaluation shows that Marabou (in divide-and-conquer mode
running on four cores) generally outperforms both Planet \cite{Planet} and
Reluplex and is able to verify all ACAS Xu benchmarks within a one hour
timeout per benchmark. 

\mbox{}

\textbf{Bastani et al.} \cite{Bastani1} presented an approach for finding
the nearest (according to the Chebyshev or $L_\infty$ distance) adversarial
example \cite{adversarial}, i.e., the closest input that causes the neural
network to produce a wrong output.
They support neural networks with piecewise linear layers (i.e.,
fully-connected, max-pooling, and convolutional layers) and ReLU
activations.
For scalability, they restrict the search to a convex region around the
given input in which the neural network is a linear function by fixing the
activation status of the ReLU activations according to the activation
status determined by the given input \cite{Montufar}. The problem is then
encoded as a linear program and fed to a black-box linear programming
solver.
The approach is demonstrated on the LeNet neural network \cite{LeNet2}
(modified to use ReLU activations instead of sigmoid activations) trained
on the MNIST dataset \cite{MNIST}, and on the larger Network-in-Network
neural network \cite{NIN} trained on the CIFAR-10 dataset \cite{CIFAR-10}.

\mbox{}

Huang et al. \cite{DLV} proposed a complementary approach for proving local
robustness to adversarial perturbations, i.e., proving that no adversarial
example exists in a neighborhood of a given input.
Their approach applies to feed-forward and convolutional neural networks
and is not tailored to specific activation functions.
It reduces the (infinite) neighborhood of a given input to a finite set of
points and checks that all these points lead to the same neural network
output. Specifically, the approach proceeds layer by layer through the
neural network, propagating constraints that relate representative points
between neural network layers.
The approach is implemented in an open-source tool named \textbf{DLV}
(\url{https://github.com/verideep/dlv}), which builds on the SMT solver
\textsc{z3} \cite{Z3}. The experimental evaluation on state-of-the-art
neural networks trained on the MNIST \cite{MNIST}, CIFAR-10
\cite{CIFAR-10}, GTSRB \cite{GTSRB}, and ImageNet \cite{ImageNet} datasets
shows that adversarial examples can be sometimes found in seconds but the
verification has prohibitive complexity for large images.

\mbox{}

Finally, a number of approaches focus on binarized neural networks
\cite{BNN}, which have been proposed as a memory efficient alternative to
traditional feed-forward neural networks. \textbf{Narodytska et al.}
\cite{Narodytska} focus on proving local robustness to adversarial
perturbations and equivalence between neural network models.
Their approach reduces verification to SAT solving, starting with a simple
MILP encoding which is first refined into an integer linear program and
then further refined into the final SAT encoding. A counterexample-guided
search procedure is then used to find property violations, or otherwise
conclude that the desired safety property is satisfied.
The experimental evaluation on neural networks trained on the MNIST dataset
\cite{MNIST} shows that the approach is able to scale to neural networks
with hundreds of neurons.

Another SAT-based approach for safety verification of binarized neural
networks was concurrently proposed by \textbf{Cheng et al.} \cite{Cheng2}.
Their approach first encodes the verification problem by means of a
combinational miter \cite{Brayton}, which is a hardware circuit with only
one Boolean output that should always be zero. The combinational miter is
then transformed into a SAT problem using standard transformation
techniques. A number of optimizations are additionally proposed to speed up
the SAT-solving time.
The optimizations allow the approach to scale to larger binarized neural
networks with thousands of neurons trained on the MNIST \cite{MNIST} and
GTSRB \cite{GTSRB} datasets.

\paragraph{MILP-based Formal Methods.}

MILP-based formal methods transform the safety verification problem into a
mixed integer linear program. Multiple different encodings have been
proposed in the literature.
For instance, consider again a $n$-layer single-output feed-forward neural
network with ReLU activations after each hidden layer, and a safety
property specifying a bounded input domain and requiring the output of the
neural network to be a positive value. A possible MILP encoding is
\cite{MIPVerify}:
\begin{subequations}
\begin{align}
	\label{eq:milp-inputA}
	&\mathbf{l} \leq \mathbf{x}_0 \leq \mathbf{u} \\
	\label{eq:milp-layerA}
	&\mathbf{\hat{x}}_{i+1} 
	= \mathbf{W}_{i+1} \mathbf{x}_i + \mathbf{b}_{i+1}
	& \forall i \in \{0, \dots, n-1\}\\
	\label{eq:milp-reluA}
	& \bm{\delta}_i \in \{0, 1\}^{|\mathbf{x}_i|},
	\quad 0 \leq \mathbf{x}_i \leq \mathbf{u}_i \cdot \bm{\delta}_i,
	\quad \mathbf{\hat{x}}_i \leq \mathbf{x}_i \leq \mathbf{\hat{x}}_i 
	- \mathbf{l}_i \cdot (1 - \bm{\delta}_i)
	& \forall i \in \{1, \dots, n-1\}\\
	\label{eq:milp-outputA}
	&min~\mathbf{x}_n
\end{align}
\end{subequations}
where Equation (\ref{eq:milp-inputA}) represents the constraints on the inputs of the neural network, 
Equation (\ref{eq:milp-layerA}) encodes the affine transformations performed
by the neural network layers, and Equation (\ref{eq:milp-reluA}) encodes the
ReLU activation functions using binary variables to represent their
activation status. Equation (\ref{eq:milp-outputA}) is the objective function, which in this case is to minimize the output of the neural network. 
A solution of this MILP problem with a negative neural network output $\mathbf{x}_n$ represents a valid counterexample. 
Otherwise, the safety property is verified.
The lower and upper bound estimations $\mathbf{l}_i$ and $\mathbf{u}_i$
for $\mathbf{x}_i$ can be obtained using interval arithmetic
\cite{IntervalArithmetic} or more precise symbolic propagation methods
\cite{ReluVal,Jiangchao,DeepPoly,Neurify}. 

\mbox{}

\textbf{Cheng et al.} \cite{Cheng1} proposed an alternative MILP formulation
which uses a variant of the Big M \cite{bigM} encoding
method for the ReLU activations:
\begin{subequations}
\setcounter{equation}{1}
\begin{align}
	\label{eq:milp-layerB}
	&\mathbf{\hat{x}}_{i+1} 
	= \mathbf{W}_{i+1} \mathbf{x}_i + \mathbf{b}_{i+1}
	& \forall i \in \{0, \dots, n-1\}\\
	\label{eq:milp-reluB}
	& \bm{\delta}_i \in \{0, 1\}^{|\mathbf{x}_i|},
	\quad 0 \leq \mathbf{x}_i \leq \mathbf{M}_i \cdot \bm{\delta}_i,
	\quad \mathbf{\hat{x}}_i \leq \mathbf{x}_i \leq \mathbf{\hat{x}}_i 
	- \mathbf{M}_i \cdot (1 - \bm{\delta}_i)
	& \forall i \in \{1, \dots, n-1\}
\end{align}
\end{subequations}
where $\mathbf{M}_i = max\{-\mathbf{l}_i, \mathbf{u}_i\}$. This encoding is
fundamentally the same as the one of Equations
(\ref{eq:milp-layerA})--(\ref{eq:milp-reluA}), except that Equation
(\ref{eq:milp-reluB}) uses symmetric bounds which are slightly worse than
the asymmetric bounds used in Equation (\ref{eq:milp-reluA}).
They address the problem of finding a lower bound on global robustness to
adversarial perturbations \cite{adversarial} with respect to the Manhattan
or $L_1$ distance, i.e., determining the largest neighborhood in which no
adversarial example exists for any possible input of a given neural network
classifier.
Specifically, they reduce the problem to finding the smallest perturbation
such that there exists an input classified as $c$ and the computed
probability $\alpha$ for $c$ may not be among the $k$ highest after a
larger perturbation. The robustness lower bound for the neural network is
the smallest among the lower bounds found for each class.
They support neural networks with piecewise linear layers (i.e.,
fully-connected, max- and average-pooling, and convolutional layers) and
ReLU and $tan^{-1}$ activations.
The $tan^{-1}$ activations are approximated piecewise using results from
the digital signal processing literature \cite{Rajan,Ukil}.
A number of heuristics are proposed to solve the resulting MILP problem
efficiently.
The approach is demonstrated on neural networks with at most $60$ inputs
and $70$ hidden neurons. The experimental evaluation shows that the
scalability of the approach relies on sometimes setting a high value for
$\alpha$.

\mbox{}

\textbf{Fischetti and Jo} \cite{Fischetti} focus on finding neural network
inputs that maximize the activation of some hidden neurons \cite{Erhan} and
building adversarial examples.
Their approach supports neural networks with piecewise linear layers (i.e.,
fully-connected, max- and average-pooling, and convolutional layers) and
ReLU activations.
They use the MILP encoding with asymmetric bounds of Equations
(\ref{eq:milp-layerA})--(\ref{eq:milp-reluA}). In order to obtain tight
lower and upper bounds $\mathbf{l}_i$ and $\mathbf{u}_i$, they proceed
layer by layer and use the MILP solver to minimize and maximize the value
of each neuron in the neural network.
The approach is demonstrated on neural networks with up to $70$ hidden
neurons trained on the MNIST dataset \cite{MNIST}, for which adversarial
examples can be found in a few minutes. For larger neural networks, the
computation time becomes too large (e.g., one hour or more for neural
networks with hundreds of neurons).

\mbox{}

Bunel et al. \cite{BaB} presented a unifying safety verification framework
named \textbf{BaB} (Branch-and-Bound) for neural networks with
piecewise-linear layers and ReLU activations. 
They show that their framework encompasses previous approaches such as
Planet \cite{Planet} and Reluplex \cite{Reluplex}.
They propose three variants of the framework: (1) BaB-relusplit, which
branches over the activation status of the ReLU activations; (2) BaB-input,
which branches over the inputs of the neural network, splitting in half
along the largest dimension; and (3) BaBSB, which performs a smarter
branching by splitting the input domain along the dimension improving the
objective function the most.
The experimental evaluation shows no particular difference between the
BaB-relusplit and BaB-input variants, while BaBSB outperforms the other
variants.
Compared to Planet and Reluplex, they perform better on shallow neural
networks, while on deeper neural networks such as the ACAS Xu networks
\cite{ACASXu} BaBSB reaches the same success rate as Reluplex but its
runtime is two orders of magnitude smaller.

\mbox{}

Dutta et al. \cite{Sherlock} proposed an approach to establish ranges for
the outputs of feed-forward fully-connected neural networks with ReLU
activations.
They discuss how to extend their approach to other activation functions by
means of piecewise linear approximations.
The approach reduces the problem to a MILP problem, using an encoding
similar to Equations (\ref{eq:milp-layerB})--(\ref{eq:milp-reluB}), and uses
local search to speed up the MILP solver.
Specifically, the approach iterates between relatively expensive MILP
solver calls and inexpensive local search iterations. The MILP solver is
used to find a solution that is better than the current target lower and
upper bounds on the output. Local search uses gradient descent
(resp.~ascent) over the neural network to find another input with a lower
(resp.~higher) lower bound (resp.~upper bound) in the input space that activates
the same ReLU activations. If such an input is found, the MILP solver is
invoked again with updated target lower and upper bounds, and so on until
the MILP solver concludes that no better solution exists.
The approach is implemented in a tool named \textbf{Sherlock}
(\url{https://github.com/souradeep-111/sherlock}), which builds on the
commercial MILP solver \textsc{Gurobi} \cite{Gurobi}.
The experimental evaluation on neural networks with thousands of neurons
shows that Sherlock outperforms a monolithic MILP approach (but requires
over three days of running time in one case). On smaller instances the
monolithic approach is faster but the running time remains small (in the
order of seconds) for both approaches. Moreover, both the monolithic
approach and Sherlock are shown to outperform Reluplex \cite{Reluplex}.

\mbox{}

\looseness=-1
Finally, Tjeng et al. \cite{MIPVerify} focus on finding the nearest
adversarial example with respect to a given distance metric, such as the
$L_1$, $L_2$, or $L_\infty$ distances.
They support neural networks with piecewise linear layers and ReLU
activations, encoded using the MILP formulation of Equations
(\ref{eq:milp-layerA})--(\ref{eq:milp-reluA}).
They use a progressive bound tightening procedure to determine lower and
upper bounds $\mathbf{l}_i$ and $\mathbf{u}_i$ for each $\mathbf{x}_i$:
they obtain initial coarse bounds quickly using interval arithmetic
\cite{IntervalArithmetic} and, only if these bounds are not sufficient to
determine the activation status of the current ReLU activation, they resort
to a more expensive procedure which uses a linear programming solver to
find tighter lower and upper bounds (by minimizing and maximizing the
value of the current neuron, as done by Fischetti and Jo \cite{Fischetti}).
The approach is implemented in a tool named \textbf{MIPVerify}
(\url{https://github.com/vtjeng/MIPVerify.jl})
and is demonstrated on neural networks with thousands of neurons
trained on the MNIST \cite{MNIST} and CIFAR-10 \cite{CIFAR-10} datasets.
MIPVerify is two to three orders of magnitude faster than Reluplex
\cite{Reluplex} in finding the nearest adversarial example with respect to
the $L_\infty$ distance.
The experimental evaluation additionally shows that the computation time
strongly correlates with the number of ReLU activations that cannot be
proven to have a fixed activation status, rather than with the total number
of ReLUs of a neural network. 

\paragraph{Other Complete Formal Methods.}

Ruan et al. \cite{DeepGO} presented an approach based on global
optimization for verifying Lipschitz continuous neural networks
\cite{adversarial}, such as neural network with piecewise linear layers
(i.e., fully-connected, max- and average-pooling, and convolutional layers)
and ReLU as well as sigmoid and tanh activations.
Specifically, given inputs to the neural networks and a Lipschitz
continuous functions over its outputs (e.g., a function determining the
confidence of the neural network in classifying an input), their approach
computes lower and upper bounds on the function values.
The approach is implemented in a tool named \textbf{DeepGO}
(\url{https://github.com/trustAI/DeepGO}).
Compared to Reluplex \cite{Reluplex} and Sherlock \cite{Sherlock}, DeepGO
on average is $36$ times faster than Sherlock and almost $100$ times faster
than Reluplex. Moreover, DeepGO is shown to be able to scale to
state-of-the-art neural networks with millions of neurons trained on the
MNIST dataset \cite{MNIST}.

\mbox{}

\begin{figure}[ht]
	\centering
	\begin{subfigure}[b]{0.45\textwidth}
		\includegraphics[width=\textwidth]{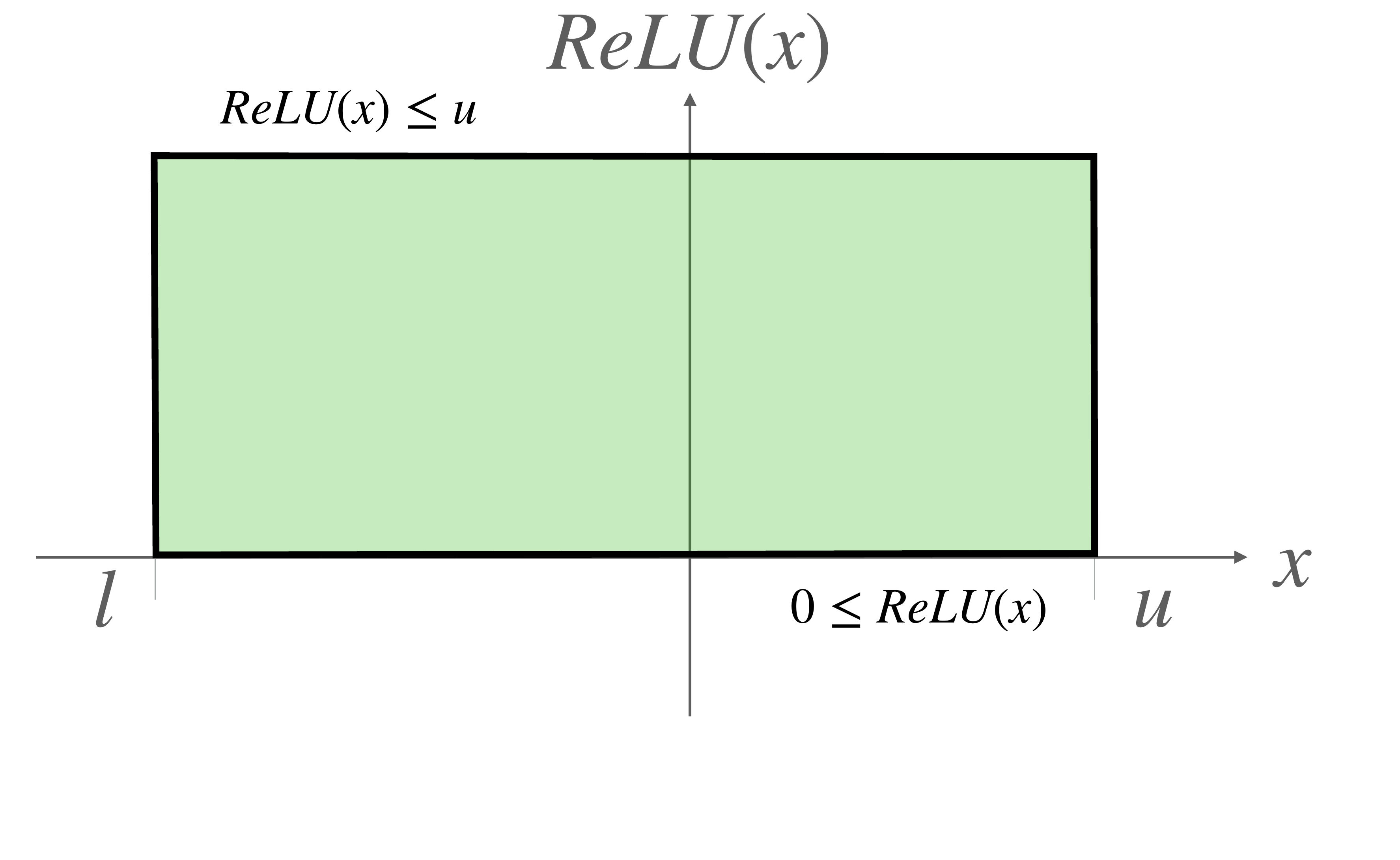}
		\caption{}
		\label{fig:naive}
	\end{subfigure}%
	\begin{subfigure}[b]{0.45\textwidth}
		\includegraphics[width=\textwidth]{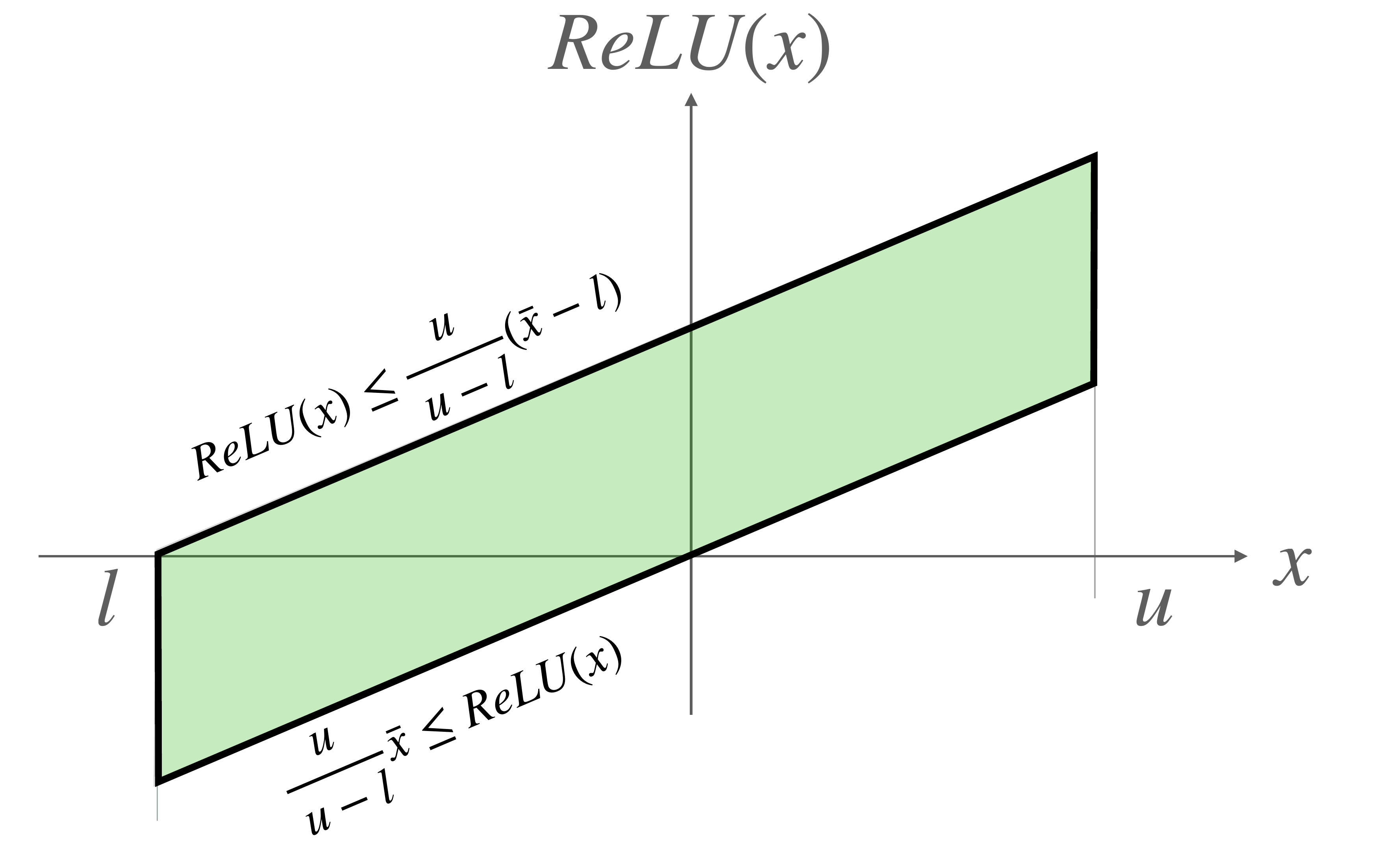}
		\caption{}
		\label{fig:neurify}
	\end{subfigure}
	\caption{Naive and Symbolic Convex Approximations 
	of a ReLU Activation.}
\end{figure}

Other complete verification methods are obtained by making incomplete
methods asymptotically complete.
For instance, the approach proposed by Wang et al. \cite{ReluVal}
iteratively refines the analyzed input space of a neural network when the
underlying over-approximating analysis is inconclusive.
They support feed-forward fully-connected neural networks with ReLU
activations.
More precisely, the over-approxi\-mating analysis uses symbolic intervals to
track lower and upper bounds for each neuron expressed whenever possible
as linear combinations of the input neurons. When the activation status of
a ReLU activation is undetermined, its range is over-approximated using the
naive convex approximation shown in Figure~\ref{fig:naive}.
When the estimated output range of the analyzed neural network is too large
to be conclusive, the input space is split and the analysis is repeated on
the resulting smaller input spaces. In particular, the input space is split
in half along the dimension that influences the most the output of the
neural network (i.e., the dimension with the largest output gradient).
Such refinement process converges in finite steps as the Lipschitz
continuity of the neural networks \cite{adversarial} ensures that the
overestimation error of the analysis decreases as the width of the input
space becomes smaller. Convergence of the approach is faster for lower
values of Lipschitz constant.
The approach is available and implemented in a tool named \textbf{ReluVal}
(\url{https://github.com/tcwangshiqi-columbia/ReluVal}). 
The experimental evaluation shows that ReluVal is on average $200$ times
faster than Reluplex \cite{Reluplex} at verifying the ACAS Xu neural
networks \cite{ACASXu}. 
According to the evaluation of Katz et al. \cite{Marabou}, ReluVal also
outperforms Marabou (in divide-and-conquer mode) when running on four
cores, while Marabou (again in divide-and-conquer mode) on average
outperforms ReluVal when running on $64$ cores.

In follow-up work, Wang et al. \cite{Neurify} presented an improved
verification tool named \textbf{Neurify}
(\url{https://github.com/tcwangshiqi-columbia/Neurify}), which supports
feed-forward and convolutional neural networks with ReLU activations.
Specifically, they improve the underlying over-approximating analysis by
partially retaining dependencies with the input neurons even when the
activation status of a ReLU activation is undetermined. Instead of using
the naive convex approximation of Figure~\ref{fig:naive}, they use the
symbolic approximation shown in Figure~\ref{fig:neurify}, where $\bar{x}$
is the symbolic lower and upper bound representation of $x$.
Additionally, Neurify iteratively minimizes the errors introduced by the
approximations by splitting on the over-approximated ReLU activations with
the larger output gradient.
On the ACAS Xu neural networks, Neurify is on average $20$ times faster
than ReluVal and $5000$ times faster than Reluplex.
The experimental evaluation also shows that Neurify is able to verify
safety properties of neural networks with over ten thousands hidden neurons
such as the self-driving car convolutional neural network Dave \cite{Dave}.

\mbox{}

Finally, other approaches offer combinations of exact and approximate
analysis. \textbf{Tran et al.} \cite{Polyhedron} used unions of bounded
convex polyhedra to compute the reachable outputs of a feed-forward
fully-connected neural network with ReLU activations.
The approach proceeds layer by layer, splitting polyhedra into two when
ReLU activations have an undetermined activation status.
Once the number of polyhedra exceeds a user-defined upper bound, the
polyhedra are clustered (using k-means clustering \cite{k-means} based on
whether they overlap) and each cluster is then over-approximated by a
hyper-rectangle.
The analysis then continues by over-approximating (again with a
hyper-rectangle) the output range of ReLU activations with an undetermined
activation status.
The precision of the approach can be improved by first partitioning the
input space and then proceeding with the analysis on each partition
independently.
The approach is demonstrated on a few easy ACAS Xu benchmarks \cite{ACASXu}
(for which the analysis requires between four minutes and over an hour),
and neural networks trained on the MNIST dataset \cite{MNIST} (for which
local robustness can be proven in under $10$ minutes for networks with
hundreds of neurons).

In follow-up work, instead of 
polyhedra, Tran et al. \cite{StarSets} used \textbf{star sets} \cite{star},
which are an equivalent representation offering fast affine mapping
operations and inexpensive intersections with half-spaces and emptiness
checking.
Their exact approach maintains a union of star sets, while their
approximate analysis computes a single star set over-approximating the
reachable outputs of a given neural network. In the latter case, ReLU
activations with an undetermined activation status are over-approximated as
shown in Figure~\ref{fig:triangle}.
The experimental evaluation on the ACAS Xu neural networks shows that their
exact analysis is on average almost $19$ times faster than Reluplex
\cite{Reluplex} and over $70$ times faster than their polyhedra-based
approach, while their approximate analysis is on average about $118$ times
faster than Reluplex.

More recently, Tran et al. \cite{ImageStars} have proposed an extension of
star sets called \textbf{image stars} to represent infinite families of
images.
They use image stars to prove local robustness to pixel brightening
perturbations \cite{AI2} of neural networks with piecewise linear layers
(i.e., fully-connected, max- and average-pooling, and convolutional layers)
and ReLU activations.
The scalability of the approach is evaluated on state-of-the-art image
classifiers such as the VGG16 and VGG19 neural networks \cite{VGG}. The
evaluation shows that the size of the analyzed input space, rather than the
size of the neural network, is the factor that mostly affects the
performance of the approach.

The polyhedra-based approach as well as the star sets-based and image
star-based approaches are all implemented in an open-source tool named
\textbf{NNV} \cite{NNV} (\url{https://github.com/verivital/nnv/}) written
in \textsc{Matlab}.
Several practical software engineering improvements have been proposed for
the exact star sets-based approach by Bak et al. \cite{nnenum} and
implemented in a tool named \textbf{NNENUM} \cite{nnenum}
(\url{https://github.com/stanleybak/nnenum}) written instead in
\textsc{Python}.
NNENUM is able to verify all ACAS Xu benchmarks in under $10$ minutes on a
standard laptop.

In currently unpublished work, \textbf{Yang et al.} \cite{FaceLattices}
proposed another alternative approach to the polyhedra-based approach of
Tran et al. \cite{Polyhedron}.
Specifically, they represent a polyhedra using its face lattice, which is a
data structure containing all the faces of a polytope (i.e., its intersections
with supporting hyperplanes, that is, its vertices, edges, ridges, and
facets) ordered by containment. 
This allows performing affine transformations, intersections with
hyperplanes, and projections on hyperplanes more efficiently than with
polyhedra.
The experimental evaluation on the ACAS Xu benchmarks shows that the face
lattice-based approach outperforms Reluplex \cite{Reluplex}, Marabou
\cite{Marabou}, and the star sets-based exact approach. ReluVal
\cite{ReluVal} instead is slightly faster overall, requiring a little over
$8$ hours to complete all benchmarks instead of almost $9$ hours needed by
the face lattice-based approach.

\mbox{}

Singh et al. \cite{RefineZono} proposed \textbf{RefineZono}, a combination
of abstract interpretation and (mixed integer) linear programming for
proving local robustness to adversarial perturbations \cite{adversarial} of
neural networks with piecewise-linear layers (i.e., fully-connected,
max-pooling, and convolutional layers) and ReLU activations.
Their approach uses the abstract domain of zonotopes \cite{ZonotopeDomain}
equipped with specialized transformers to over-approximate the neural
network activations \cite{DeepZ}. Specifically, each neuron of a neural
network is associated with a zonotope and concrete lower and upper bounds
over-approximating its possible values.
In order to determine lower and upper bounds, for scalability, they use
faster but less precise methods as the analysis proceeds to deeper layers
of the neural networks: for the first user-defined number of layers, they
rely on mixed integer linear programming as done by Fischetti and Jo
\cite{Fischetti}; for the next user-defined number of layers, they
approximate ReLU activations as in Figure~\ref{fig:triangle} and use a
linear programming solver to find possibly coarser bounds; for the
remaining layers, they simply continue the analysis with the abstract
domain without additional bound tightening.
When the analysis is inconclusive (i.e., the over-approximation of the
neural network outputs is too large), the collected lower and upper bounds
can be used to encode the robustness verification as a MILP problem (thus
making the approach complete).
The experimental evaluation on a neural network with $150$ neurons trained
on the MNIST dataset \cite{MNIST} shows that RefineZono is faster than
MIPVerify \cite{MIPVerify} for complete verification of local robustness to
perturbations with respect to the $L_\infty$ distance (up to $\epsilon =
0.03$). The approach is also demonstrated on one of the ACAS Xu benchmarks
\cite{ACASXu} known to be hard: by first uniformly splitting the input
space into $6300$ regions, RefineZono is able to verify the benchmark over
four times faster than Neurify \cite{Neurify}.

\subsubsubsection{Incomplete Formal Methods}
\label{sec:incomplete}

Incomplete formal verification methods are generally \emph{able to scale}
to large neural network architectures (they generally require at most a few
minutes for neural networks with thousands of neurons) and are \emph{sound}
(often also with respect to floating-point arithmetic
\cite{DeepZ,DeepPoly,Jiangchao}) but suffer from \emph{false positives}.
Incomplete methods are also generally less limited to certain neural
network architectures and activations (e.g., some also apply to recurrent neural networks \cite{POPQORN,Zhang2}).
We distinguish below between methods based on \emph{abstract
interpretation} \cite[etc.]{AI2,DeepZ,DeepPoly,Jiangchao,OOPSLA20}, and
other incomplete methods based on simulation \cite{MaxSens}, duality \cite{WongKolter1,Dvijotham}, semidefinite programming
\cite{Raghunathan}, or linear approximations \cite{Fast-Lin}.

\paragraph{Abstract Interpretation-based Methods.}

The first use of abstract interpretation for verifying neural networks was
the \textbf{AI}$^{\mathbf{2}}$ framework proposed by Gehr et al.
\cite{AI2}. They address the problem of proving local robustness to
adversarial perturbations \cite{adversarial} for neural networks with
piecewise linear layers (i.e., fully-connected, max-pooling, and
convolutional layers) and ReLU activations.
Their approach over-approximates the computations performed by a neural
network by interpreting them in a chosen abstract domain. Thus, starting
from an abstraction of the inputs of the neural network, it propagates the
abstraction forward through the network layers taking affine transformations
and activation functions into account, until it reaches the output layer
where the abstraction represents an over-approximation of the neural
network outputs.
Their framework supports the abstract domains of intervals
\cite{IntervalDomain}, zonotopes \cite{ZonotopeDomain}, and polyhedra
\cite{PolyhedraDomain}, as well as their bounded powersets.
In the experimental evaluation they considered pixel brightening
perturbations (up to $\delta = 0.085$) against image classifiers (with up
to $53000$ neurons \cite{LeNet1}) trained on the MNIST \cite{MNIST} and and
CIFAR-10 \cite{CIFAR-10} datasets.
The results show that the domain of zonotopes offers a good trade-off
between precision and scalability of the analysis. It takes on average less
than $10$ seconds to verify local robustness for feed-forward neural
networks with up to $18000$ hidden neurons trained on MNIST. On the largest
convolutional neural networks trained on CIFAR-10 however the difference in
precision with respect to the less expressive domain of intervals is less
significant.

In a follow-up work, Singh et al. \cite{DeepZ} presented a custom zonotope
abstract domain named \textbf{DeepZ} equipped with specially designed
abstract transformers for ReLU as well as sigmoid and tanh activations. 
The transformers are made sound with respect to floating-point arithmetic
using the methodology proposed by Miné \cite{Mine}.
In the experimental evaluation they consider perturbations with respect to
the $L_\infty$ distance (up to $\epsilon = 0.3$) against image classifiers
(with up to $88500$ hidden neurons) again trained on MNIST and CIFAR-10. The results show that DeepZ is significantly more precise and
faster than AI$^2$. Moreover, DeepZ is able to verify local robustness
properties of the largest convolutional neural network within a few minutes.

\begin{figure}[ht]
	\centering
	\begin{subfigure}[b]{0.45\textwidth}
		\includegraphics[width=\textwidth]{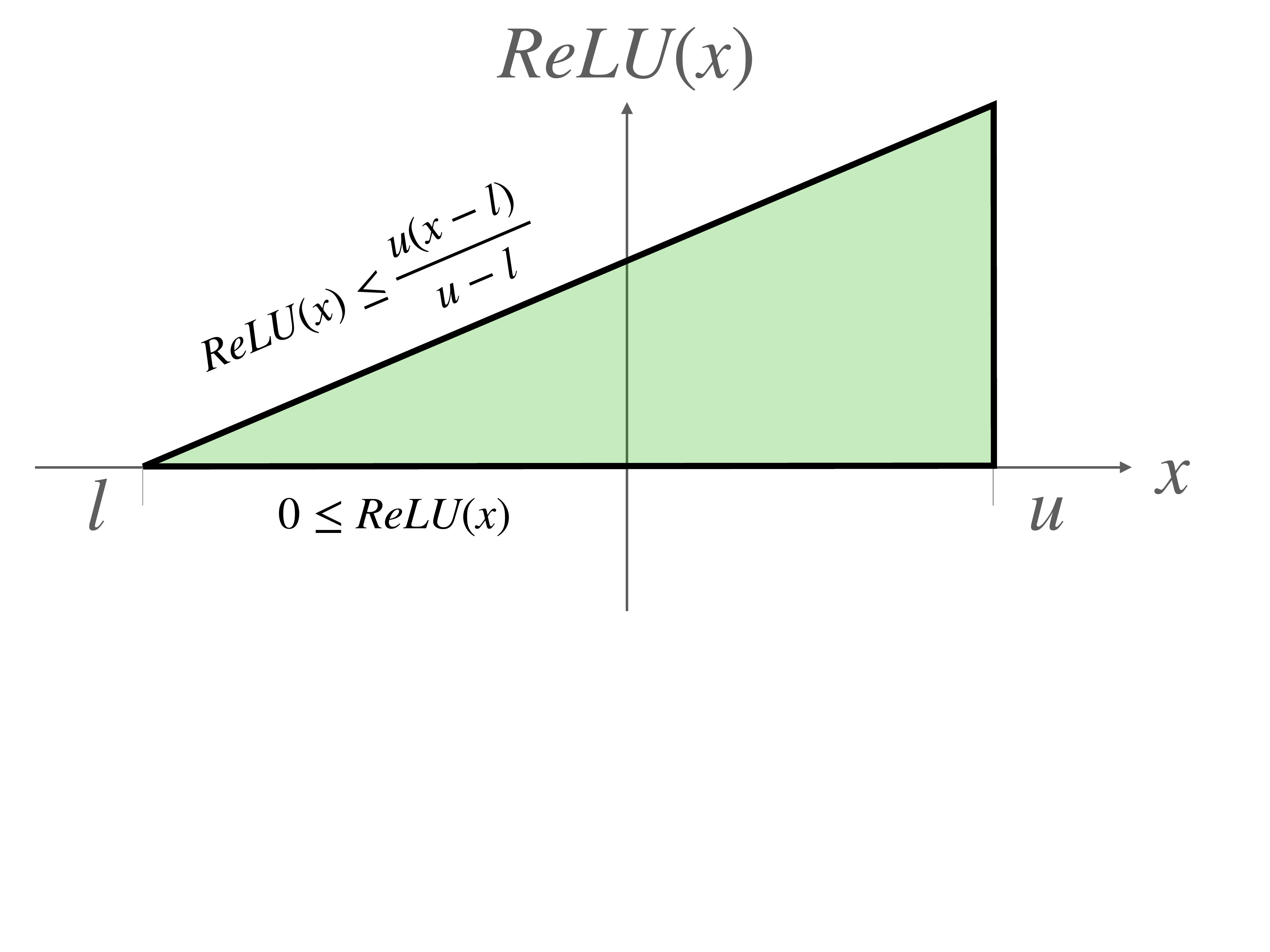}
		\caption{}
		\label{fig:deeppoly1}
	\end{subfigure}%
	\begin{subfigure}[b]{0.45\textwidth}
		\includegraphics[width=\textwidth]{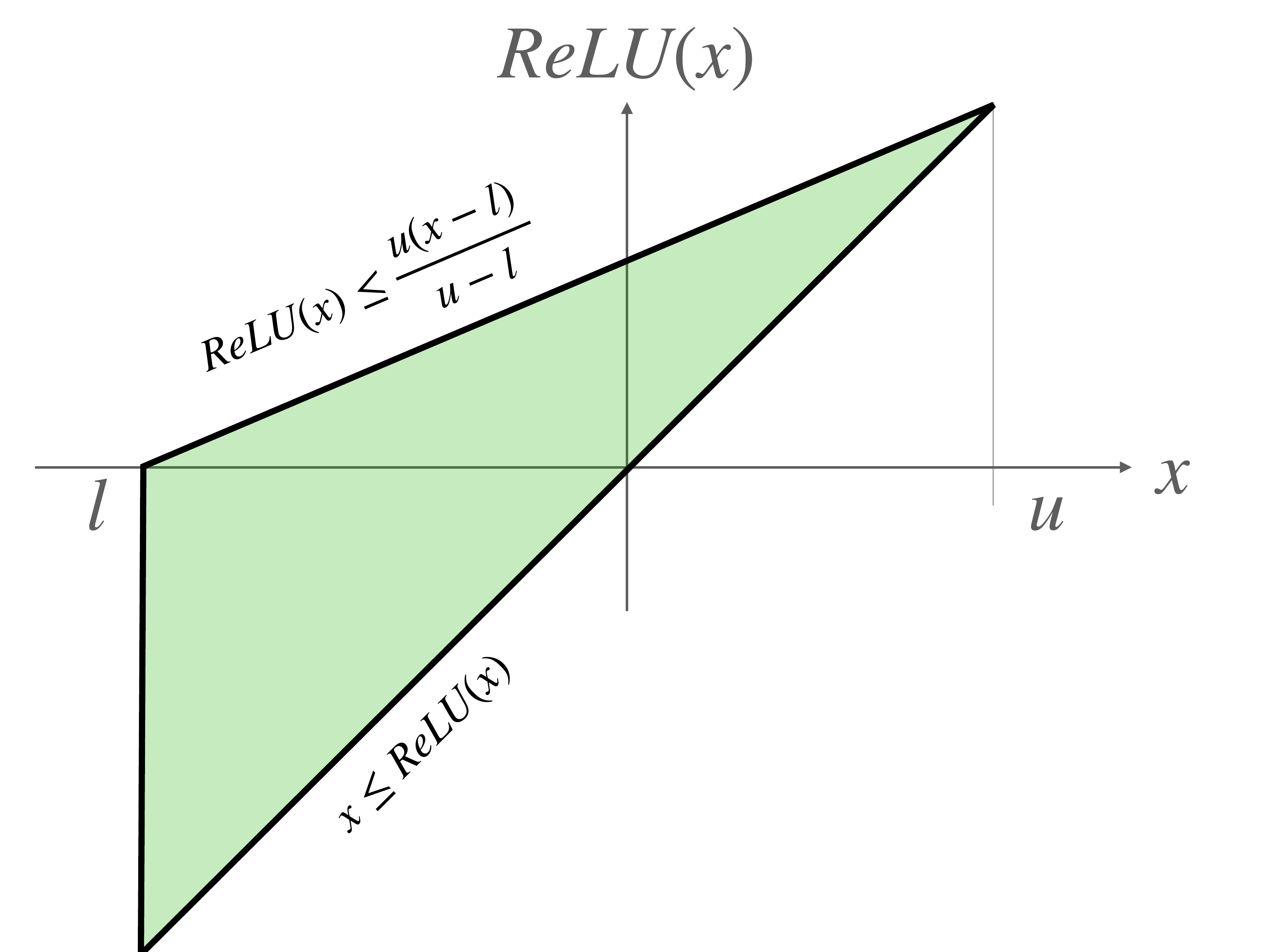}
		\caption{}
		\label{fig:deeppoly2}
	\end{subfigure}
	\caption{Alternative Convex Approximations 
	of a ReLU Activation.}\label{fig:deeppoly}
\end{figure}

Another abstract domain proposed by Singh et al. \cite{DeepPoly} is the
\textbf{DeepPoly} domain.
DeepPoly associates to each neuron of a neural network concrete lower and
upper bounds as well as symbolic bounds expressed as linear combinations of
the neurons in the preceding layer of the network.
It is additionally equipped with abstract transformers specifically
designed for ReLU, sigmoid, and tanh activations.
In particular, the ReLU abstract transformer first computes concrete lower
and upper bounds $l$ and $u$ for a given neuron $x$ by back-substitution of
its symbolic bounds up to the input layer. Then, it chooses the tighter
approximation for $ReLU(x)$ between those in Figure~\ref{fig:deeppoly},
i.e., the approximation in Figure~\ref{fig:deeppoly1} when $u \leq -l$, and
that in Figure~\ref{fig:deeppoly2} otherwise.
The transformers are again made sound with respect to floating-point
arithmetic using the methodology proposed by Miné \cite{Mine}.
In the experimental evaluation they consider image classifiers (with up to
$88500$ hidden neurons) trained on MNIST and CIFAR-10, and perturbations
with respect to the $L_\infty$ distance (up to $\epsilon = 0.3$) as well as
rotations (between $-45^{\circ}$ and $65^{\circ}$).
The results show that DeepPoly is generally even faster and often strictly
more precise than DeepZ.

Singh et al. \cite{k-ReLU} later presented the \textbf{k-ReLU} framework to
jointly approximate $k$ ReLU activations in order to produce a more precise
over-approximation of a neural network layer.
They instantiate the framework with the DeepPoly domain resulting in a new
abstract domain called kPoly.
The experimental evaluation considers perturbations with respect to the
$L_\infty$ distance (up to $\epsilon = 0.3$) against image classifiers
(with up to $107496$ neurons) trained on MNIST and CIFAR-10.
The results show that kPoly is more precise than DeepPoly as well as
RefineZono \cite{RefineZono}. kPoly is also faster than RefineZono and has
an average runtime of less than $8$ minutes.

DeepZ, DeepPoly, RefineZono, and kPoly are all implemented in a tool named
\textbf{ERAN} (\url{https://github.com/eth-sri/eran}), which builds on the
ELINA abstract domain library \cite{ELINA} and the commercial MILP solver
\textsc{Gurobi} \cite{Gurobi}.

In currently unpublished work, \textbf{Müller et al.} \cite{GPUPoly} have
presented algorithms for running DeepPoly efficiently on a GPU. The
experimental evaluation shows that this makes the analysis up to $170$
times faster than DeepPoly and able to scale to neural networks with up to
$967000$ neurons trained on the CIFAR-10 dataset.

\mbox{}

\textbf{Li et al.} \cite{Jiangchao} proposed a symbolic propagation
technique to improve the precision of abstract interpretation-based
verification methods for neural networks with piecewise-linear layers
(i.e., fully-connected, max-pooling, and convolutional layers) and ReLU
activations. Specifically, the value of each neuron in the neural network
is represented symbolically as a linear combination of the input neurons
and the values of the ReLU activations in previous layers. In particular,
ReLU activations with an undetermined activation status are represented by
fresh symbolic variables.
Their approach generalizes the symbolic propagation technique used in
ReluVal \cite{ReluVal} to any abstract domain.
In the experimental evaluation, they combine their symbolic propagation
technique with the abstract domains of intervals \cite{IntervalDomain} and
zonotopes \cite{ZonotopeDomain}, and focus on verifying local robustness to
perturbations with respect to the $L_\infty$ distance (up to $\epsilon =
0.6$) against image classifiers (with up to $89572$ neurons) trained on the
MNIST \cite{MNIST} and CIFAR-10 \cite{CIFAR-10} datasets.
The results show that their symbolic propagation technique significantly
improves precision with a modest running time increase (it is less than two
times slower in most cases and often faster).

\mbox{}

More recently, Urban et al. \cite{OOPSLA20} presented an approach for
verifying fairness of feed-forward neural networks with ReLU activations
used for classification of tabular data. The fairness notion that they
consider is a form of global robustness, which requires that the output
classification of a neural network is not affected by the values of the
neural network inputs that are determined to be sensitive to bias by the
user \cite{Galhotra}.
The approach combines a forward analysis with an inexpensive abstract
domain (e.g., the domain of intervals \cite{IntervalDomain} possibly
combined with the symbolic propagation technique proposed by Li et al.
\cite{Jiangchao}, or the DeepPoly domain \cite{DeepPoly}) to group together
the partitions of the neural network input space that share the same ReLU
activation pattern, and a more expensive backward analysis with the domain
of polyhedra \cite{PolyhedraDomain}, to determine the partitions that are
subject to bias.
The approach is implemented in an open-source tool named \textbf{Libra}
(\url{https://github.com/caterinaurban/Libra}) and evaluated on neural
networks with up to over a thousand neurons trained on the Adult
\cite{Adult}, COMPAS \cite{COMPAS}, German Credit \cite{German}, and
Japanese Credit \cite{Japanese} datasets.
The results show that the performance of the approach correlates with the
size of the analyzed input space rather than the size of the analyzed
neural network (as also observed by Tran et al. \cite{ImageStars}). The
analysis might require a few days to complete on the full input space of a
neural network with $20$ hidden neurons, but only about three hours on a
fraction of the input space of a neural network with $1280$ hidden neurons.

\mbox{}

Another complementary line of work focuses on abstracting a given neural
network by computing a smaller network which over-approximates the outputs
of the original network.
\textbf{Prabhakar and Afzal} \cite{INN} propose an abstraction for
feed-forward neural networks with ReLU activations. Their construction
merges neurons layer-wise according to a partitioning strategy. The weights
of the neural network are carefully replaced with intervals
\cite{IntervalDomain} to account for the merging, which ensures that the
outputs of the abstraction soundly over-approximate the outputs of the
original neural networks.
The output range analysis on the resulting abstract neural network is
reduced to solving a MILP problem, using an encoding similar to Equations
(\ref{eq:milp-layerB})--(\ref{eq:milp-reluB}) extended to deal with interval
weights.
The approach is demonstrated on the ACAS Xu neural networks \cite{ACASXu},
experimenting with abstractions with different numbers of abstract neurons
(up to $32$ abstract neurons). 
The results show that the abstraction considerably reduces the computation
time needed for the analysis. On the other hand, the precision of the
analysis is affected by the choice of the partitioning strategy used to
guide the merging of neurons in the abstraction.

The construction of Prabhakar and Afzal was recently generalized by
\textbf{Sotoudeh and Thakur} \cite{Sotoudeh} to other (convex) abstract
domains than the domain of intervals and other activations than ReLUs (e.g.,
Leaky ReLU \cite{LeakyReLU} as well as sigmoid and tanh activations).
They identify and formally justify sufficient and necessary conditions on
the abstract domains and activations to guarantee soundness and, when
these conditions are not satisfied, they propose workarounds to modify the
original neural network into an equivalent one that satisfies the
conditions.

\mbox{}

\textbf{Elboher et al.} \cite{Elboher} recently presented instead an
abstraction of feed-forward neural networks with ReLU activations based on
counterexample-guided abstraction refinement (CEGAR) \cite{CEGAR}.
Their approach first transforms a neural network into an equivalent network
in which each neuron belongs to one of four classes, determined by the
weights of its outgoing connections (i.e., whether the weights are all
negative or positive) and its effect on the neural network output (i.e.,
whether increasing its value increases or decreases the network output).
A basic abstraction step then merges two neurons that belong to the same
class, while a basic refinement steps splits previously merged neurons. The
choice of which neurons to merge or split is done heuristically and several
possible heuristics are proposed.
The experimental evaluation on $90$ of the ACAS Xu benchmarks shows that
Marabou \cite{Marabou} combined with their abstraction framework
outperforms Marabou alone. Within a $20$ hours timeout, it is able to
verify $58$ benchmarks, while Marabou alone can only verify $35$
benchmarks. The abstraction-enhanced Marabou is also orders of magnitude
faster than Marabou alone.

\mbox{}

Finally, a couple of theoretical results on the abstract
interpretation of neural networks can be found in the literature.
\textbf{Baader et al.} \cite{UAT} show that for any continuous function $f$
there exists a feed-forward fully-connected neural network with ReLU
activations whose abstract interpretation using the domains of intervals
\cite{IntervalDomain} from an input region $B$ is an arbitrarily close
approximation of $f$ on $B$. 
In particular, their result reduces to the classical universal
approximation theorem \cite{Leshno} when the input region $B$ is a single
input point.
In practice this means that, for any neural network, there exists another
arbitrarily close neural network that can be more easily verified.
The result trivially holds for more expressive abstract domains than
intervals.

In currently unpublished work, \textbf{Wang et al.} \cite{AUA} generalize
the result of Baader et al. beyond ReLU activations (e.g., including
sigmoid and tanh activations).

\paragraph{Other Incomplete Formal Methods.}

Among other incomplete verification methods, \textbf{Xiang et al.}
\cite{MaxSens} proposed an approach that combines simulation and linear
programming to over-approximate the outputs of feed-forward neural networks
with monotonic activations (e.g., ReLU as well as sigmoid and tanh
activations).
The approach first discretizes the input space of the neural network into
$L_\infty$ balls centered around individual inputs. The ball radius is
user-defined and determines the granularity of the discretization.
Then, simulations are executed for each ball center to determine the
corresponding neural network outputs.
Finally, the approach proceeds layer by layer to measure the maximum
deviation of the layer output with respect to the other points in each
$L_\infty$ ball by solving a finite set of convex optimization problems. At
the output layer this yields an over-approximation of the neural network
outputs.
The finer the initial input space discretization (i.e., the smaller the
user-defined ball radius), the tighter the resulting over-approximation.
The approach is demonstrated on a small use case and the results show that
it scales well with respect to the number of hidden layers of a neural
network but poorly with respect to the size of its input space (i.e., the
number of input dimensions).

\mbox{}

Other incomplete methods focus on providing robustness guarantees to
adversarial perturbations \cite{adversarial} with respect to a given
distance metric, i.e., finding a non-trivial lower bound on the distance to
the nearest adversarial example for any training input (or, equivalently,
determining the largest neighborhood for any training input in which no
adversarial example exists).
The approach of \textbf{Wong and Kolter} \cite{WongKolter1} is based on
duality and supports neural networks with piecewise linear layers (i.e.,
fully-connected, max-pooling, and convolutional layers) and ReLU
activations, and perturbations with respect to any $L_p$ distance.
Specifically, similarly to Ehlers \cite{Planet}, they reduce the problem to
a linear program by approximating ReLU activations as in
Figure~\ref{fig:triangle}. However, since the resulting formulation has a
number of variables equal to the number of ReLUs in the neural network,
they instead solve its dual formulation, for which any feasible solution
gives a lower bound on the solution of the primal.
The approach is demonstrated on relatively small neural networks (e.g.,
networks with two fully-connected and two convolutional layers trained on
MNIST \cite{MNIST} and Fashion-MNIST \cite{Fashion-MNIST}), and the authors
highlight the difficulty of scaling-up the approach to larger neural
network architectures such as ImageNet classifiers \cite{ImageNet}.
According to the evaluation of Tjeng et al. \cite{MIPVerify}, the gap
between the lower bound found by Wong and Kolter and the minimum
adversarial perturbation found by MIPVerify is significant even for small
neural networks, and increases for larger network architectures.

\mbox{}

Concurrently to the work of Wong and Kolter, \textbf{Raghunathan et al.}
\cite{Raghunathan} reduce the problem to solving a single semidefinite
program for each pair of output labels of a neural network (rather than
for each training input as Wong and Kolter). Their approach however only
supports feed-forward fully-connected neural networks with one hidden
layer, and perturbations with respect to the Chebyshev or $L_\infty$
distance.

\mbox{}

\textbf{Dvijotham et al.} \cite{Dvijotham} presented another approach based
on duality for verifying safety properties of feed-forward neural networks
with arbitrary activations. 
They formulate the verification problem as global optimization problem
that seeks to find the largest violation to the desired safety property. If
the largest violation is smaller than zero, the safety property is
satisfied.
To circumvent the need to solve a non-convex optimization, they reduce the
problem to finding an upper bound to the largest violation, which they do
by solving the dual of a Lagrangian relaxation \cite{Lagrangian} of the
original optimization problem.
Interestingly, for feed-forward neural networks with ReLU activations, this
dual is the dual of the linear programming formulation of Ehlers
\cite{Planet} (but it is different from the dual derived by Wong and Kolter
\cite{WongKolter1}).
The approach is compared to that of Raghunathan et al. \cite{Raghunathan}
and is shown to compute better bounds on robustness to $L_\infty$
perturbations on neural networks trained on MNIST \cite{MNIST}.

\mbox{}

Other approaches for computing robustness lower bounds are the
\textbf{Fast-Lin} and \textbf{Fast-Lip} algorithms proposed by Weng et al.
\cite{Fast-Lin} for feed-forward fully-connected neural networks with ReLU
activations.
The Fast-Lin algorithm is based on the direct computation of layer-wise
lower and upper bounds for each neuron using symbolic linear
approximations similarly to Neurify \cite{Neurify} (cf.
Figure~\ref{fig:neurify}), while the Fast-Lip algorithm is based on
bounding the neural network local Lipschitz constant.
The experimental evaluation (on neural networks with hundreds of neurons
trained on MNIST \cite{MNIST}) shows that the lower bounds against
$L_\infty$ perturbations found by Fast-Lin and Fast-Lip are only two to
three times larger than the minimum adversarial perturbation found using
Reluplex \cite{Reluplex}, but Fast-Lin and Fast-Lip are over $10000$ times
faster.
Fast-Lin and Fast-Lip are also shown to be able to scale to and find
non-trivial robustness lower bounds for larger neural network with
thousands of neurons.

In follow-up work, Zhang et al. \cite{CROWN} have proposed a generalization
of Fast-Lin to feed-forward fully-connected neural networks with arbitrary
activations (e.g., ReLU, sigmoid, tanh, and arctan activations).
The approach is available and implemented in a tool named \textbf{CROWN}
(\url{https://github.com/CROWN-Robustness/Crown}).
The experimental evaluation (on neural networks with thousands of neurons
trained on MNIST \cite{MNIST} and CIFAR-10 \cite{CIFAR-10}) shows that
CROWN finds lower bounds that are between $19\%$ and $20\%$ better than
those found by Fast-Lin at the cost of a modest increase in computation
time (CROWN being less than two times slower than Fast-Lin).

More recently, Boopathy et al. \cite{CNN-Cert} presented a further
generalization of CROWN named \textbf{CNN-Cert}
(\url{https://github.com/AkhilanB/CNN-Cert}) to support convolutional
neural networks, and Ko et al. \cite{POPQORN} presented \textbf{POPQORN}
(\url{https://github.com/ZhaoyangLyu/POPQORN}) which supports recurrent
neural networks, including long short-term memory \cite{LSTM} and gated
recurrent unit \cite{GRU} architectures.

\mbox{}

Another recent approach targeting recurrent neural networks was proposed by
\textbf{Zhang et al.} \cite{Zhang2}. They focus on proving safety properties
of vanilla recurrent neural networks with ReLU activations trained for
cognitive tasks \cite{Song}.
Their approach consists in first training an equally performing but easier
to verify neural network using idea from Xiao et al. \cite{Xiao}, i.e.,
using regularization during training to encourage the stabilization of ReLU
activations.
Then, they unfold the trained network for the number of time steps on which
the desired safety property is defined, and proceed layer by layer to
compute the reachable outputs of the network using unions of polytopes.
Specifically, they encode each polytope and input-output relation of a
layer as a MILP problem and use a solver to partition the polytope into
sub-polytopes that can each be linearly mapped to a convex polytope in the
next layer.
Since the number of polytopes increases with the number of layers, they
propose two techniques to keep the representation tractable.
A first option is a CEGAR-based \cite{CEGAR} technique that joins polytopes
when their number exceeds a user-controlled threshold, and backtraces to
refine the abstraction when the resulting over-approximation is too large
to be conclusive. In particular, they join polytopes that originate from
the same polytope in the previous neural network layer.
Another option is to attempt to find an inductive invariant, that is, a
polytope such that its image through a neural network layer is contained in
the polytope. This requires a fixpoint iteration with some form of widening
(i.e., iteration acceleration) to ensure termination.
The approach is demonstrated on the verification of a number of properties
of a single use case. Their invariant generation technique overall appears
to be the most precise albeit not the fastest one.

\mbox{}

Finally, \textbf{Gopinath et al.} \cite{Gopinath} presented an approach for
inferring safety properties of feed-forward neural networks with ReLU
activations. 
Specifically, they infer properties that correspond to activation patterns
of the network ReLU activations. Input properties are convex predicates on
the network input space that imply the desired output behavior, and
correspond to activation patterns that constrain the activation status of
some ReLUs in the neural network. Layer properties instead correspond to
activation patterns that only constrain the activation status of ReLUs at
some hidden layer. They thus express unions of convex regions in the input
space that imply the desired output behavior.
To infer input properties, they start from an input that satisfies the
given postcondition and iteratively relax the corresponding activation
pattern in order to constrain the activation status of the least number of
ReLUs but still satisfy the given postcondition. 
Instead, to directly learn layer properties, they train a decision tree on input-output pairs and their
corresponding activation pattern.
The approach is demonstrated on one of the ACAS Xu neural networks
\cite{ACASXu} and two neural networks (with hundreds of neurons) trained on
the MNIST dataset \cite{MNIST}.

\subsubsection{Formal Methods for Other Machine Learning Models}
\label{sec:ml}

We discuss below formal verification methods for other machine learning models than neural networks. We distinguish between methods for \emph{support vector machines} (Section~\ref{sec:svm}) and methods for 
\emph{decision tree ensembles} (Section~\ref{sec:ensembles}).

\subsubsubsection{Formal Methods for Support Vector Machines}
\label{sec:svm}

\begin{figure}[ht]
	\centering
	\includegraphics[width=0.75\textwidth]{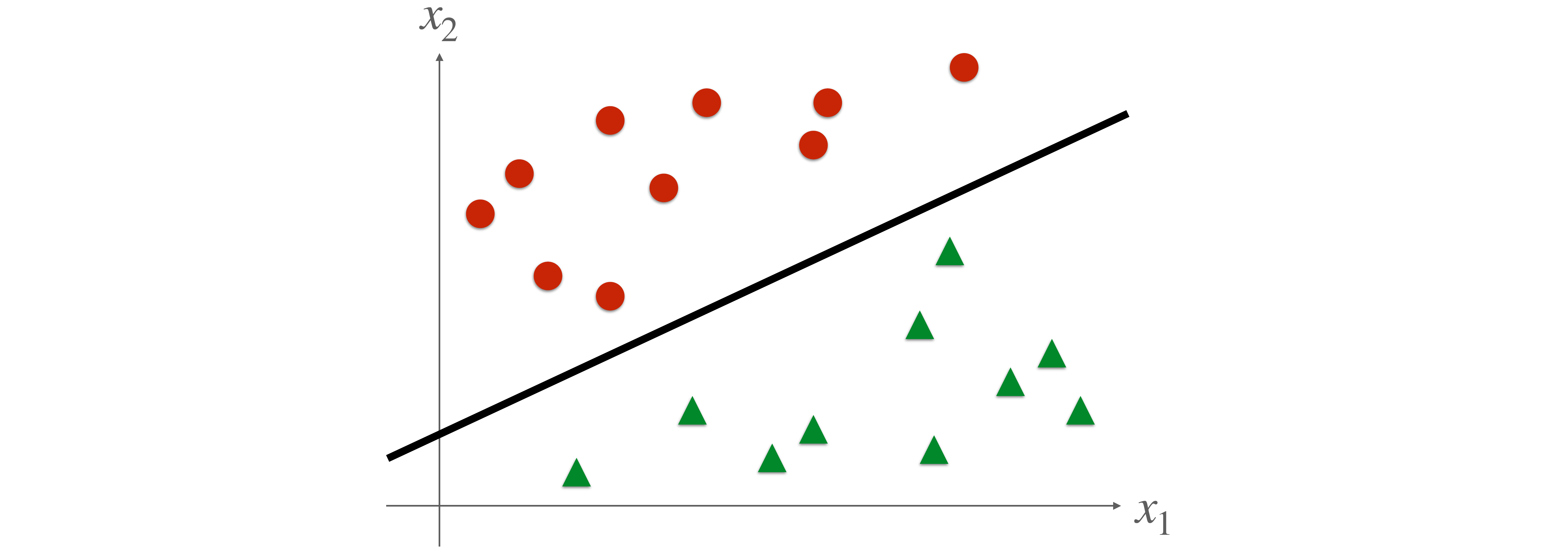}
	\caption{Linear Kernel Support Vector Machine.}\label{fig:svm}
\end{figure}

\noindent
Support vector machines \cite{SVM} partition their input space in regions
using \emph{hyperplanes} that separate the training inputs according to
their labels.
They hyperplanes are chosen (i.e., learned during training) to maximize
their distance (called \emph{margin}) to their closest training inputs
(called \emph{support vectors}).
Figure~\ref{fig:svm} shows a very simple support vector machine in a
two-dimensional input space.
When the training inputs are not linearly separable in the input space, the
input space is implicitly projected, using so-called \emph{kernel
functions}, into a much higher-dimensional space in which the training
inputs become linearly separable. This allows a support vector machine to
also learn non-linear decision boundaries.

\mbox{}

To the best of our knowledge, the only verification method for support
vector machines was proposed by Ranzato and Zanella \cite{SAVer}.
They focus on proving local robustness to adversarial perturbations of
support vector machines based on the most commonly used kernel functions
(i.e., linear, polynomial, and radial basis function kernels \cite{SVM}).
Their approach is based on abstract interpretation and uses the abstract
domain of intervals \cite{IntervalDomain} combined with reduced affine
forms \cite{RAF}, which is a restriction of the domain of zonotopes
\cite{ZonotopeDomain} to affine forms of a given length (i.e., the
dimension of the input space of the analyzed support vector machine). In
particular, reduced affine forms capture dependencies between input
dimensions.
The approach is implemented and available in an open-source tool named
\textbf{SAVer} (\url{https://github.com/abstract-machine-learning/saver}).

\subsubsubsection{Formal Methods for Decision Tree Ensembles}
\label{sec:ensembles}

\vspace{-1em}
\begin{figure}[ht]
	\centering
	\includegraphics[width=0.75\textwidth]{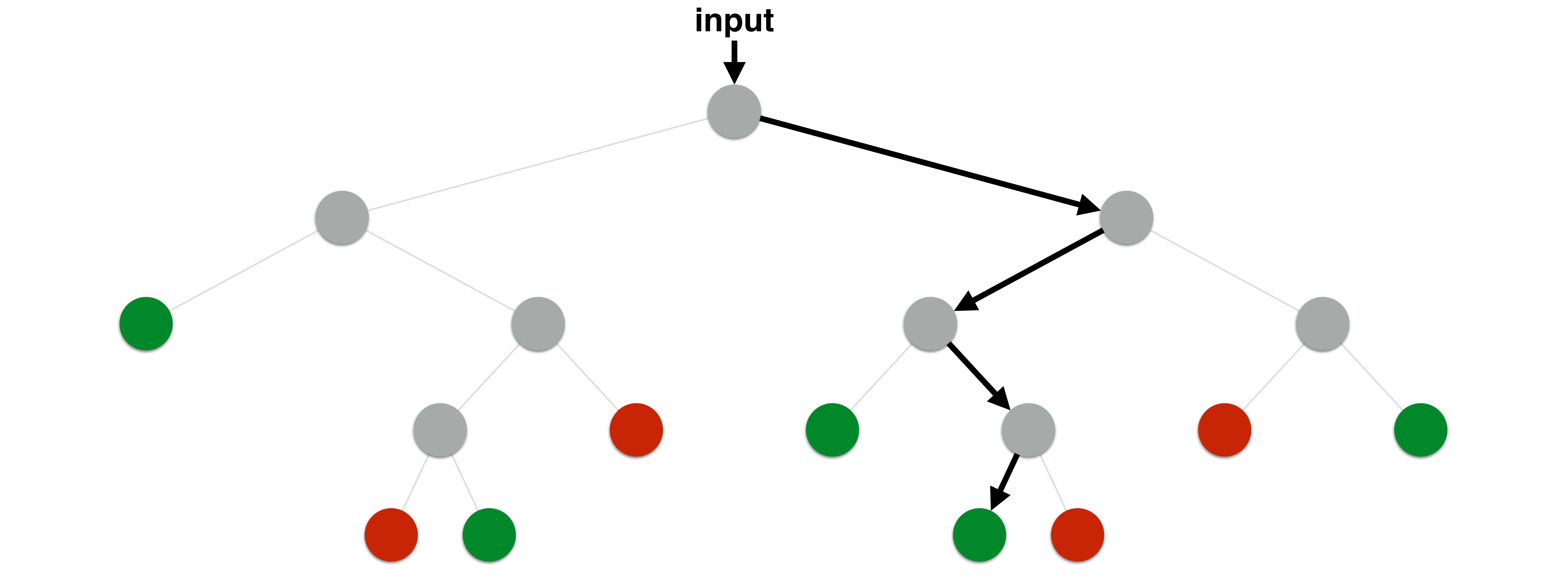}
	\caption{Decision Tree for Binary Classification.}\label{fig:tree}
\end{figure}

\noindent
Decision trees \cite{CART} are used for both classification and regression
tasks. The leaves of \emph{classification trees} are labeled with a class
(or a probability distribution over the classes skewed towards a particular
class), while the leaves of \emph{regression trees} can take continuous
values (typically real numbers).
The internal nodes of a decision tree recursively partition the input
space. Each node is labeled with a Boolean split criterion over the input
dimensions that decides whether to branch to the left or right subtree.
Most often, decision trees employ univariate splits over a single input
dimension with respect to some value.
Thus, the prediction for a given a input is determined starting at the root
of the decision tree and recursively branching to the appropriate subtree
according to the encountered split criteria, until ultimately a leaf node
is reached that decides the prediction (cf. Figure~\ref{fig:tree}).

Decision tree ensembles are sets of decision trees which together
contribute to formulate a unique classification or regression output.
One of the most used decision tree ensembles are \emph{random forests}
\cite{RF}, where each tree of the ensemble is trained independently from
the other trees on some random subset of the input dimensions. The final
output is typically obtained through a voting mechanism (e.g., majority
voting).
Another popular decision tree ensemble are \emph{gradient boosted decision
trees} \cite{GBDT}, where the ensemble is built incrementally by training
each new tree on the basis of the input data samples that are misclassified
by the previous trees.

\mbox{}

Among the earliest verification methods for decision tree ensembles
\textbf{Kantchelian et al.} \cite{Kantchelian} proposed an approach for
finding the nearest adversarial example with respect to the $L_0$, $L_1$,
$L_2$, and $L_\infty$ distances.
They focus on sum-ensembles of regression trees with univariate split
predicates (i.e., the prediction of the ensemble is the sum of all
individual tree predictions) used for binary classification (i.e., by
thresholding of the prediction of the ensemble).
They reduce the problem to solving a MILP problem. The MILP encoding is
linear in the size of the tree ensemble but the solving time can be
prohibitive for ensembles that consist of a high number of trees or with a
large number of input dimensions. The approach is demonstrated on decision
tree ensembles trained on the MNIST dataset \cite{MNIST}.

\mbox{}

More recently, \textbf{Chen et al.} \cite{Chen2} presented a linear time
algorithm for finding the nearest adversarial for a decision tree. They
support decision trees with univariate split predicates and box-shaped
perturbations around a given input. In particular, they focus on
perturbations with respect to the $L_\infty$ distance.
Their approach consists in performing a depth-first traversal of the
decision tree to compute a box for each of its leaves such that all inputs
in a box fall into the corresponding leaf. Then, the minimal adversarial
example corresponds to the minimal perturbation required to change a given
input to be into a different box (and thus fall into a different leaf).
For decision tree ensembles, they cast the problem into a max-clique
searching problem in $K$-partite graphs, where $K$ is the number of trees
in the ensemble. In the case of large decision tree ensembles (i.e., with
thousands of trees), they propose an algorithm that supports any-time
termination and provides a lower bound on local robustness.
The experimental evaluation compares the latter approach to that of
Kantchelian et al. \cite{Kantchelian} and shows that it gives tight lower
bounds with a speed-up of over $3000$ times in one case (i.e., an ensemble
with $300$ trees trained on the HIGGS dataset \cite{LIBSVM}).

\mbox{}

\textbf{Sato et al.} \cite{Sato} proposed an SMT-based approach for safety
verification of random forests and gradient boosted decision trees.
Specifically, their approach focuses on finding inputs that lead to a
violation of a given output property.
They reduce the verification to solving an SMT encoding of the problem.
They additionally propose a technique for generalizing a single input
counterexample to a range of inputs violating the output property.
The approach is demonstrated on the verification of three properties of an
ensemble with $100$ trees with maximum depth $3$. Further experiments on
scalability show a strong correlation between the size of the ensemble and
the verification time. In particular, the verification time increases by
$10$ to $100$ times when the depth of the trees is increased by one.
Overall, the approach is practical for ensembles with less than $200$ trees
with maximum depth less than $5$.

\mbox{}

Another SMT-based approach was previously presented by \textbf{Einziger et
al.} \cite{VeriGB} for verifying local robustness to adversarial
perturbations of gradient boosted decision trees used for classification.
Their approach focuses on perturbations with respect to the $L_\infty$
distance.
They encode the problem as an SMT instance and rely on an off-the-shelf SMT
solver for the verification. In order to speed-up the verification they
propose two optimizations, i.e., pruning constraints encoding decision tree
paths that are too distant from the given input, and checking local
robustness with respect to each output class in parallel.
The approach is implemented in a tool named VeriGB, built on top of the SMT
solver \textsc{Z3} \cite{Z3}.
The experimental evaluation on ensembles trained on the MNIST dataset
\cite{MNIST} shows the scalability limitations of the approach (which often
requires over $10$ minutes for the verification of ensembles with $100$
trees with depth up to $10$). Interestingly, the evaluation also shows that
ensembles with similar accuracy but lower tree depth tend to be more
robust. Indeed, it has been observed that trees with high depth suffer more
from over-fitting \cite{Hastie}.

\mbox{}

Other approaches for proving local robustness to adversarial perturbations
are (more or less explicitly) based on abstract interpretation.
Törnblom and Nadjm-Tehrani proposed an abstraction-refinement approach
for random forests \cite{VoRF} and gradient boosted decision trees
\cite{VoTE1,VoTE2} with univariate split predicates.
The approach iteratively partitions the
neighborhood of perturbations around a given input into hyper-rectangles
until the analysis becomes conclusive (i.e., it is able to prove local
robustness or it has found a real counterexample).
It is implemented in an open-source tool named \textbf{VoTE}
(\url{https://github.com/john-tornblom/VoTE}).

\mbox{}

A more general framework that supports arbitrary abstract domains was
presented by Ranzato and Zanella \cite{silva}. They instantiate the
framework with the domain of intervals \cite{IntervalDomain} and show that
the approach is sound and complete for decision tree ensembles with
univariate split predicates.
The approach is available and implemented in an open-source tool named
\textbf{Silva} (\url{https://github.com/abstract-machine-learning/silva}).
The experimental evaluation on ensembles (with up to $150$ trees with depth
$10$) trained on the MNIST dataset \cite{MNIST} shows that Silva is
generally faster than VoTE \cite{VoRF,VoTE1,VoTE2} and able to scale to
larger models.

\mbox{}

Finally, very recently, \textbf{Calzavara et al.} \cite{TreeCert} have
proposed an approach that supports perturbations modeled as arbitrary
imperative programs (as opposed to distance-based perturbations).
Their approach consists in first performing a straightforward conversion of
a decision tree and an input perturbation into a loop-free imperative
program, and then analyzing the resulting program using abstract
interpretation (e.g., with the domain of polyhedra \cite{PolyhedraDomain}).
The approach generalizes to decision tree ensembles, but it is not complete
(unlike VoTE \cite{VoRF,VoTE1,VoTE2} and Silva \cite{silva}).

\subsection{Formal Methods for Earlier Pipeline Phases}
\label{sec:other}

We conclude the overview of the state of the art by discussing formal
methods that apply to earlier phases in the machine learning pipeline (cf. Figure~\ref{fig:pipeline}).
We distinguish between methods that apply to the \emph{model training}
phase (Section~\ref{sec:training}) and methods that apply to the \emph{data
preparation} phase (Section~\ref{sec:preparation}).

\subsubsection{Formal Methods for Model Training}
\label{sec:training}

Formal methods for model training proposed so far in the literature have a
rather narrow application scope, focusing on \emph{robust training} to
obtain machine learning models that are more robust against adversarial
input perturbations.
The problem of adversarial perturbations has been initially addressed in
the context of spam email detection \cite{Dalvi,Lowd}. For neural networks,
the problem was brought up by the seminal work of Szegedy et al.
\cite{adversarial}.
Robust training aims at minimizing the worst-case loss for each given
input, that is, the maximum loss (i.e., model error) over all possible
perturbations of a given training input. However, since calculating the
exact worst-case loss can be computationally costly in practice, robust
training approaches typically minimize an estimate of the worst-case loss.

\mbox{}

\emph{Adversarial training} approaches minimize a lower bound on the
worst-case loss. Specifically, they rely on fast ways to generate
adversarial inputs under a considered threat model
\cite[etc.]{Goodfellow,DeepFool,Huang} and use them to augment the training
data \cite[etc.]{Goodfellow,Huang,Kurakin}.
Instances of this method are typically empirically shown to be successful
against known adversaries. A notable result for neural networks is the work
of Madry et al. \cite{Madry} against the strong adversarial attack of
Carlini and Wagner \cite{Carlini}.
Other approaches include, for instance, the work of Kantchelian et al.
\cite{Kantchelian}, Calzavara et al. \cite{Calzavara}, and Chen et al.
\cite{Chen1} for decision tree ensembles.
While promising, these approaches do not guarantee that models are trained
to be robust against any kind of adversarial perturbation around an input
under the considered threat model.

\mbox{}

To address this lack of guarantees, \emph{certified training} approaches
instead minimize an upper bound on the worst-case loss in order to
guarantee robustness on all training inputs.
Among these are the approach of \textbf{Andriushchenko and Hein}
\cite{Andriushchenko} for decision trees, and the early approach of
\textbf{Hein and Andriushchenko} \cite{Hein} and the previously described
concurrent approaches of \textbf{Wong and Kolter} \cite{WongKolter1} and
\textbf{Raghunathan et al.} \cite{Raghunathan} for neural networks.
In particular, Wong and Kolter and Raghunathan et al. essentially use their
robustness lower bounds as regularizer during training to encourage
robustness against all perturbations with respect to the considered
distance metric.
The experimental evaluation conducted by Wong and Kolter, shows that the
latter approaches are complementary: Wong and Kolter achieve lower
robustness test error but higher traditional test error than Raghunathan et
al.
The approach of Wong and Kolter has been later generalized to support
neural networks with arbitrary activations \cite{WongKolter2}
However, scalability to large neural network architectures remains an
issue. \textbf{Mirman et al.} \cite{DiffAI} considerably improve in
scalability by proposing a certified training approach based on abstract
interpretation. 

\mbox{}

Overall, there is a trade-off between robustness and accuracy of a trained
model \cite{Tsipras}. The stronger robustness guarantees given by certified
training approaches come at the cost of a significantly lower accuracy
compared to models trained using adversarial training.
Recently, \textbf{Balunović and Vechev} \cite{COLT} addressed this problem
by combining adversarial and certified training and obtained
state-of-the-art robustness and accuracy results on the CIFAR-10 dataset
\cite{CIFAR-10} (with respect to $L_\infty$ perturbations with $\epsilon =
2/255$) improving over the concurrent work of Zhang et al. \cite{Zhang1}.

\mbox{}

To the best of our knowledge, addressing broader verification goals beyond
robust training (e.g., what Kurd and Kelly \cite{Kurd} identify as goal G3)
currently remains an unexplored direction in the literature.

\subsubsection{Formal Methods for Data Preparation}
\label{sec:preparation}

Software employed in earlier stages of the machine learning pipeline
to gather, triage, and pre-process ``dirty'' data is the most fragile of
the entire pipeline as it generally heavily relies on implicit assumptions
on the data.
It is also often disregarded as single-use non-critical glue code and, for
this reason, is poorly tested, let alone formally verified.
However, poor choices or accidental mistakes (e.g., programming errors)
made in this phase can have an important effect on the end result of the
pipeline and, more dangerously, can remain completely unnoticed as a
plausible end result gives no indication that something went wrong along
the way \cite{Herndon}.

\mbox{}

Formal verification methods that address these issues are scarce in the
literature. An exception is the work of \textbf{Urban and Müller}
\cite{ESOP18}, which proposed an abstract interpretation framework for
reasoning about data usage and a static analysis method for automatically
detecting (possibly accidentally) unused input data.
Other work in this space, is a currently unpublished configurable static
analysis for automatically inferring assumptions on the input data
\cite{Urban}.
Much more work is however still needed, especially in applications in which
software deals with complex unstructured data and is written by domain
experts which are not necessarily software engineers.

\section{Research Perspectives}\label{sec:future}

The overview of the state of the art shows that we are still far away from
being able to verify the \emph{entire} machine learning pipeline,
which we argue is necessary to ensure the safe use of machine learning software in safety-critical applications. We conclude by
discussing below a number of research directions that we have identified
for each pipeline phase and that will bring us closer to this objective.

\paragraph{Data Preparation.}

For data preparation software, there is a need for further approaches for
reasoning about data usage \cite{ESOP18}. For instance, verification
methods that detect (possibly accidentally) reused and duplicated data
would be a valuable complement to existing approaches. More generally,
approaches for tracking data provenance would provide important
\emph{traceability} guarantees. This is particularly useful when data is
aggregated from different sources (which is often the case).

\looseness=-1
Another direction worthy of being pursued further is the inference of
assumptions on the input data that are embedded in data preparation
software \cite{Urban}. Specifically, designing more sophisticated
abstractions would allow inferring more informative assumptions. In turn,
these assumptions could be leveraged by tools that assist in the data
preparation, e.g., by automatically generating (provably safe) data
cleaning code.

\paragraph{Model Training.}

In the context of model training, designing verification methods with
broader application scope than robust training (cf.
Section~\ref{sec:training}) is of primal importance. In particular, there
is a need for approaches providing stronger formal guarantees on the
training process. For instance, it would be useful to be able to verify how
the behavior of the model being trained evolves on a particular input space
of interest. To this end, it might be interesting to leverage the theory
behind the mathematical tools used for training to design domain-specific
precise abstractions.

Another interesting possibility is to develop approaches to determine
constraints on the training process that enforce a desired property on the
trained model. For instance, one could enforce a specific behavior for a
given input range of interest, or a particular structure of the trained
model that facilitates its subsequent verification \cite{Xiao} (or its
abstraction for verification \cite{INN,Sotoudeh,Elboher}).

\paragraph{Model Deployment.}

For trained models, future work should aim at verifying more interesting
properties, beyond what exists in the literature. For instance,
domain-specific robustness properties are not necessarily limited to
perturbations based on distance metrics or image perturbations, but might
be complex specifications over a (not necessarily convex) input space of
interest. Tackling such properties will require adaptations of
existing abstractions or even the design of completely new ones. Moreover,
verifying \emph{global} robustness properties rather than local robustness
properties would be more adapted to safety-critical applications.

More generally, verification methods should allow verifying the behavior of
trained models under \emph{all} circumstances and not just over the
expected safe input space. This is particularly useful in cases in which
the trained model is part of a larger machine learning-based system
\cite{VerifAI} and is thus more susceptible to receive unexpected inputs,
e.g., from malfunctioning sensors. Addressing such issues will likely
require designing more sophisticated verification approaches (e.g.,
possibly combining forward and backward analyses \cite{OOPSLA20}) and
overcoming scalability challenges \cite{ImageStars,OOPSLA20} (e.g., by
designing abstractions able to leverage GPUs \cite{GPUPoly}).

Another interesting case is when the trained model is approximating an
existing exact reference model \cite{ACASXu}. In this case, it would be
useful to design verification methods able to relate the behavior of the
inexact model to the optimal expected behavior, e.g., by determining the
worst-case approximation of the trained model.

When the verification fails, having verification methods able to determine
the source of the fault \cite{Deng} (e.g., a particular group of nodes in a
neural network) would, for instance, allow localizing and guiding any
repairs. More generally, such verification methods would enhance
interpretability of models such as neural networks.

Finally, there is a need for further approaches that support diverse model
structures, such as recurrent neural networks \cite{POPQORN,Zhang2}, and
that can verify model implementations by taking
floating-point rounding errors into account \cite{DeepZ,DeepPoly,Jiangchao}.

\section*{Acknowledgements}
This work is partially supported by 
Airbus and 
the European Research Council under Consolidator Grant Agreement 681393 --- MOPSA.

\bibliographystyle{plain}


\end{document}

%% file: abstract.tex
\begin{abstract}
  We review state-of-the-art formal methods applied to the emerging field of the verification of machine learning systems.
  Formal methods can provide rigorous correctness guarantees on hardware and software systems.
  Thanks to the availability of mature tools, their use is well established in the industry, and in particular to check safety-critical applications as they undergo a stringent certification process.
  As machine learning is becoming more popular, machine-learned components are now considered for inclusion in critical systems.
  This raises the question of their safety and their verification.
  Yet, established formal methods are limited to classic, i.e. non machine-learned software.
  Applying formal methods to verify systems that include machine learning has only been considered recently and poses novel challenges in soundness, precision, and scalability.

  We first recall established formal methods and their current use in an exemplar safety-critical field, avionic software, with a focus on abstract interpretation based techniques as they provide a high level of scalability.
  This provides a golden standard and sets high expectations for machine learning verification.
  We then provide a comprehensive and detailed review of the formal methods developed so far for machine learning, highlighting their strengths and limitations.
  The large majority of them verify trained neural networks and employ either SMT, optimization, or abstract interpretation techniques.
  We also discuss methods for support vector machines and decision tree ensembles, as well as methods targeting training and data preparation, which are critical but often neglected aspects of machine learning.
  Finally, we offer perspectives for future research directions towards the formal verification of machine learning systems.
\end{abstract}

%% file: fm.tex

This section gives an informal overview of current formal methods for software verification, notably abstract interpretation, and selected recent examples of applications on embedded safety-critical software.
We refer to \cite{CousotMarktoberdorf} for another short introduction to abstract interpretation and to \cite{MineTutorial} for a tutorial.

\subsection{Overview of Formal Methods}
\label{sec:fm:overview}

Formal methods are an array of techniques that employ logic and mathematics in order to provide rigorous guarantees about the correctness of computer software.
The first principled formal methods date back from the pioneering work of Floyd \cite{Floyd}, Hoare \cite{Hoare} and Naur \cite{Naur} on program logic in the late 60s, although similar ideas have been attested as far back as the late 40s with a notable work by Turing \cite{Turing}.
Early works describe such methods through pen-and-paper proofs of tiny programs in idealized languages.
Developing and checking by hand the very large proofs that are required to verify real-sized applications in realistic languages would be intractable.
Hence, the last few decades have seen the development of software verification tools.

Keeping in mind the fundamental undecidability of correctness properties of programs as a consequence of Rice's Theorem \cite{Rice}, it is in fact impossible to design a tool that can decide precisely for every program whether it is correct or not.
Tools must abandon either full automation (thus requiring some manual assistance), generality (handling only a subset of programs), or completeness (sometimes reporting as incorrect a correct program, which we call a false alarm, or failing to terminate).
In the rest of the section, we discuss the broad categories of computer-assisted verification methods that have been subsequently proposed, and the trade-offs they make to solve these issues in practice.
These approaches also vary in the set of correctness properties that can be checked and how the programmer can express them.
However, a requirement of all formal methods is that they should be \emph{sound}, that is, any correctness property established by the tool is indeed true of the actual executions of the program.
In practice, tools may only be sound for subsets of programming languages (e.g., not supporting the ``eval'' construction), or for an idealized semantics (e.g., using reals instead of floating-point numbers), or by making implicit assumptions that must be checked by other means (e.g., no aliasing between pointers passed to functions) \cite{Soundiness}.
A formal verification tool must then make these limitations explicit and state clearly the formal guarantees that are expected despite the limitations.

\paragraph{Deductive Verification.}
Deductive methods stem directly from the work of Floyd, Hoare, and Naur \cite{Floyd,Hoare,Naur}.
The user provides a program and a formal specification expressed in a logic language.
The specification is then propagated through the program source code, using symbolic execution as proposed by Burstall \cite{Burstall} or weakest preconditions as proposed by Dijkstra \cite{Dijkstra}.
This generates automatically a set of verification conditions implying that the program obeys its specification.
The conditions are then proved correct with the help of a solver, such as a fully automated SMT (satisfiability modulo theory) solver (e.g., \textsc{z3} \cite{Z3}).

One benefit of deductive verification is its ability to handle complex specifications using decidable logical theories (reasoning about, e.g., integers, floating-point numbers, pointers, arrays, recursive data-structures) and to perform modular verification by breaking down the specification into function contracts with preconditions and postconditions.
However, automation is limited by the need to supplement the specification with additional annotations, notably loop variants and invariants, contracts for internal functions, and ghost variables so that the verification conditions become tractable for the prover.
The progress of solvers has made deductive verification an attractive approach, but there is still the occasional need to help the solver through interactive proofs.
Examples of current deductive verification platforms include Why 3 \cite{Why3}, as well as Frama-C \cite{FramaC} used to check industrial C software.

\paragraph{Design by Refinement.}
Design by refinement is a related approach based on successive refinements of a sequence of state machines, from an abstract specification up to an executable implementation.
Each refinement step is proven formally with the aid of an automated solver.
This technique requires a large specification effort and is suited to develop formally verified software from the ground up rather than verifying existing software.
A popular instance of this technique is the B method \cite{B}.

\paragraph{Proof Assistants.}
Interactive proof assistants are general-purpose tools able to produce reliable proofs of theorems, with application to mathematics, logic, and computer science.
Their strength lies in the use of very expressive logic and the ability to check every proof with a small trusted core, reducing the possibilities of errors.
Although they can offer some degree of automation (e.g., through tactics), they employ undecidable logics and essentially rely on a very tight interaction between the programmer and the tool, which can prove time consuming.
Unlike automated SMT solvers, the limit of what can be proved with proof assistants lies solely in the ingenuity (and time and dedication) of the user.
One popular proof assistant is Coq \cite{Coq}, which also acts as a programming language and is able to automatically extract an executable implementation from a constructive proof.
Proof assistants have been applied to prove involved correctness properties on a few complex pieces of software, such as the CompCert certified C compiler \cite{CompCert}.
In the case of CompCert, Coq was used to prove to equivalence between a source code and its compiled version (provided that the former is free of undefined behavior), which is a very strong example of functional correctness result.

\paragraph{Model Checking.}
In the 80s, Clarke and Emerson \cite{Clarke} and Queille and Sifakis \cite{Sifakis} independently invented model checking, which checks specifications on finite-state models of hardware or software systems by exhaustive exploration.
It provides a sound, complete, and automatic verification method on models.
Early model checking methods relied on explicit state representations, which suffer from combinatorial explosion and severely limit the size of the models considered.
Symbolic model checking \cite{McMillan} subsequently introduced more compact representations that also permit the verification of some regular classes of infinite-state models.

The benefits of model checking are achieving both soundness and completeness, the use of temporal logic \cite{Pnueli} to express rich specifications (including termination and other liveness properties), and the ability to check concurrent models.
Model checking has been particularly successful to check hardware systems (for instance at Intel \cite{intel}) which are inherently finite.
However, whenever the system to be checked is infinite-state, or is simply too large, it is the responsibility of the user to provide a simplified model.
Crafting a model that is both faithful and efficiently checked is a difficult task.
Even then, the soundness and completeness properties only hold with respect to the model, not the original system.

More recently, software model checking has targeted the direct verification of source code without requiring a hand-crafted model, but several trade-offs had to be made to sidestep the intractability of an exhaustive exploration of the state space.
Bounded model checking \cite{BMC} (e.g., the CBMC tool \cite{CBMC}) limits the exploration to execution traces of fixed length.
The drawback of this method is that it loses soundness and completeness with respect to the full program; while it can uncover errors occurring at the beginning of the execution of the program, it cannot prove the absence of errors.
Counterexample-guided abstraction refinement (CEGAR) \cite{CEGAR} automatically extracts an abstract model from the source code based on a finite set of predicates, which is then verified using model-checking.
If the correctness proof fails, the model checker extracts a counterexample which is used to refine automatically the abstract model, and the process is iterated until the proof is established.
The CEGAR method has been successfully used to check the correctness of Windows device drivers at Microsoft \cite{slam}.
In certain cases, however, the refinement process does not terminate.

\paragraph{Semantic Static Analysis.}
Static program analyzers perform a direct and fully automated analysis of the source code of a program without executing it.
In the broad sense, the term ``static analysis'' also includes syntactic style checkers (so-called ``linters'').
We only focus here on semantic-based static analyzers that offer formal guarantees on the result of the analysis, i.e., they belong to the category of formal methods.
To achieve full automation, termination, and scalability, static analyzers interpret programs at an abstract level, focusing only on the properties of interest.

A classic method for static analysis is data-flow analysis, introduced in the 70s by Kildall \cite{Kildall}.
This technique infers program properties by propagating abstract values from a finite-height lattice of properties along the control flow of the program.
The method is popular in compilers as it features very efficient algorithms and can infer properties useful for optimization (constant propagation, live variable analysis, etc.).
However, such properties are generally not expressive enough to support program verification.
More general analyzers can be constructed through abstract interpretation, a theory of the approximation of program semantics introduced by Cousot and Cousot \cite{CC77}.
They are not limited to finite-height abstractions and can, for instance, infer variable bounds and relationships, handle dynamic memory allocation, etc.
While less efficient than data-flow analyses used in compiler, they always terminate and remain relatively efficient thanks to abstractions that ignore irrelevant (or too complex) details and perform approximations.
They nevertheless guarantee soundness by erring on the safe side: they over-approximate the set of possible behaviors of programs when they cannot be modeled faithfully, so that any property inferred as true on the over-approximation is also true on the original program.
The benefit of static analysis by abstract interpretation is full automation, parameterized precision, and the ability to verify large programs in real languages at the source level (C, Java, etc.) or binary level.
The analysis is however inherently incomplete, and can fail to verify a desired property due to over-approximations.
Thus, it can output false alarms, that must be checked by other means.

Static analyzers by abstract interpretation can perform whole-program analysis to prove non-functional properties, such as absence of run-time errors, that are implicitly specified by the language and embedded in the choice of abstraction.
An example is the Astrée analyzer \cite{Astree} for embedded C code, discussed in the next section.
They can also perform a modular proof of user-provided contracts, as in CodeContracts \cite{CodeContracts}, achieving a similar goal to deductive verification but with full automation, by removing the need to annotate the program with loop invariants or contracts for private functions.
In each tool, the abstractions have been tailored to the class of properties to be verified.
Compared to classic model checking and deductive methods, the effort has thus shifted from the user to the analysis designer.
We discuss abstract interpretation in more details in Section~\ref{sec:fm:ai}.

\subsection{Applications in Embedded Critical Software}
\label{sec:fm:critical}

Formal methods have transitioned from academia to hardware and software industries, with many individual success stories (see \cite{FMindus} for a recent review).
We present here an account, based on \cite{Airbus,Delmas}, of its use on safety-critical embedded software in the avionics industry.
It is a significant example as this industry is an early adopter of formal methods due to the stringent safety requirements on software, although such methods have recently made their way into less critical industries such as Facebook \cite{Facebook}.
Nevertheless, it provides a picture of a fully realized integration of these methods and also hints at the level of maturity expected from formal methods for artificial intelligence to become usable in safety-critical applications.

As a critical aircraft component, avionic software is subject to certification by certification authorities and must obey the DO-178 international standard, which notably describes the verification process that such software must follow.
Traditional methods consist in massive test campaigns and intellectual reviews, which have difficulties scaling up and maintaining a reasonable cost (which already accounts for more than half of the cost of overall software development).
A shift occurred in the 2010s with the new DO-178C revision (and its DO-333 supplement) which opens the door to certification by formal methods.
One key aspect of the standard is the emphasis on employing \emph{sound formal methods}.
Another key aspect is that any verification technique (formal or not) employed at the source level cannot be considered valid at the binary level unless the compilation process is also certified, which is important in practice as some tools can only reason at the source level.

Classic certification processes combine a variety of techniques (reviews, unit tests, integration tests, etc.).
Likewise, when introducing formal methods into the verification process, Atki et al. \cite{Delmas} present an array of different tools and techniques that are currently used (or are considered in the future) to supplement or replace legacy techniques with, as goal, to improve industrial efficiency while maintaining the requested safety and reliability level:
\begin{itemize}
\item
  Deductive verification with Frama-C \cite{FramaC} is used to implement unit proofs of individual C functions of some software subsets.
  The article reports that 95\% of the proof obligations are solved automatically by the SMT solver, and the remaining part requires interactions with the prover.
  This activity can partially replace classic unit testing, with the additional benefit of ensuring that the properties tested are fully formalized in a contract language.
\item
  The Astrée \cite{Astree} abstract interpretation-based static analyzer is used to check for the absence of run-time errors as defined by the C standard (integer and floating-point arithmetic errors, pointer or array access errors, invalid operations, etc.) as well as assertion failures.
  Astrée performs a whole-program analysis at the C source level, scaling to programs of a few million lines with very few false alarms.
\item
  The Fluctuat \cite{Fluctuat} abstract interpretation-based static analyzer is used to assess the numerical accuracy of floating-point computations in C libraries of control programs.
\item
  The StackAnalyzer and aiT WCET abstract interpretation-based static analyzers from AbsInt (\url{https://www.absint.com}) are used at the binary level on the compiled code in order to compute (over-approximations of) the worst-case execution time and the worst-case stack usage, and ensure that they do not exceed the allocated resources.
\item
  The CompCert certified compiler \cite{CompCert} is used to compile the C code into binary.
  This ensures that the compilation process does not introduce any error that was absent from the source.
  Note that, although the design of CompCert required a massive proof effort in Coq, its use by avionic software programmers does not require any effort at all.
\end{itemize}
These tools are complementary.
For instance, the unit proof of Frama-C implicitly assumes the absence of invalid pointers, aliasing, or arithmetic overflows to optimize proof automation.
Likewise, CompCert assumes that the C source has no undefined behaviors and does not offer any guarantee otherwise.
These properties are ensured by the whole-program analysis performed by Astrée.
In return, CompCert ensures the equivalence of the source and binary code, hence, the properties that Frama-C and Astrée validated on the C source can be assumed to hold on the binary code.

We can conclude from the choice of formal methods reported in \cite{Delmas} that two desirable properties for formal verification methods of safety-critical software are \emph{soundness} (including with respect to floating-point computations) and \emph{automation}.

\subsection{Abstract Interpretation}
\label{sec:fm:ai}

\looseness=-1
Abstract interpretation is a general theory of the approximation of program semantics introduced in the late 70s by Cousot and Cousot  \cite{CC77}.
It provides mathematical tools to compare different semantics and prove crucial properties, such as soundness and completeness, and formalizes the notions of approximation and abstraction.

\paragraph{Static Analysis by Abstract Interpretation.}
One important practical application of abstract interpretation is the design of static analyzers.
It allows deriving, in a principled way, static analyzers from program semantics focusing on a desired program property, ensuring soundness by construction.
This is achieved by applying a sequence of abstractions until the semantics becomes effectively computable.

\begin{figure}[ht]
	\centering
	\begin{subfigure}[b]{0.32\textwidth}
		\centering
		\includegraphics[height=3.25cm]{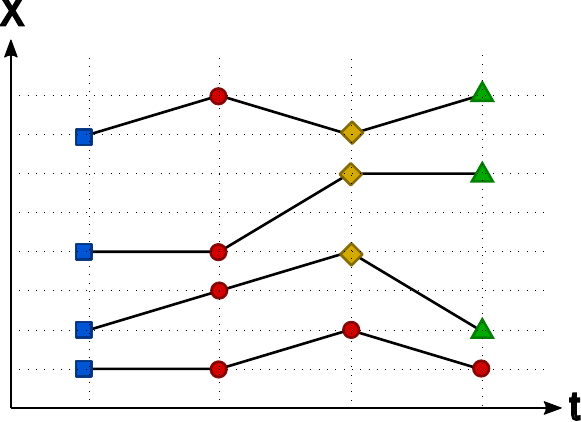}
		\caption{}\label{fig:trace-a}
	\end{subfigure}
	\begin{subfigure}[b]{0.32\textwidth}
		\centering
		\includegraphics[height=3.25cm]{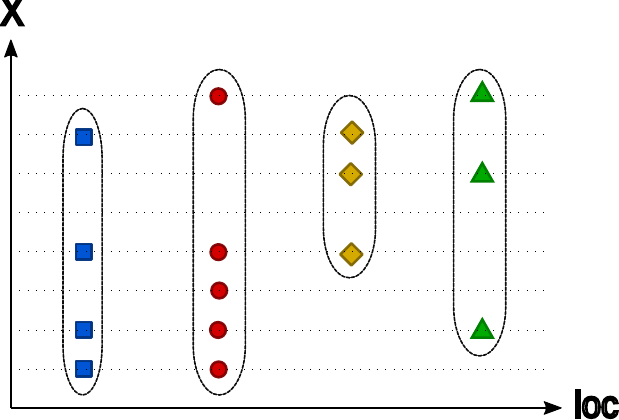}
		\caption{}\label{fig:trace-b}
	\end{subfigure}
	\begin{subfigure}[b]{0.32\textwidth}
		\centering
		\includegraphics[height=3.25cm]{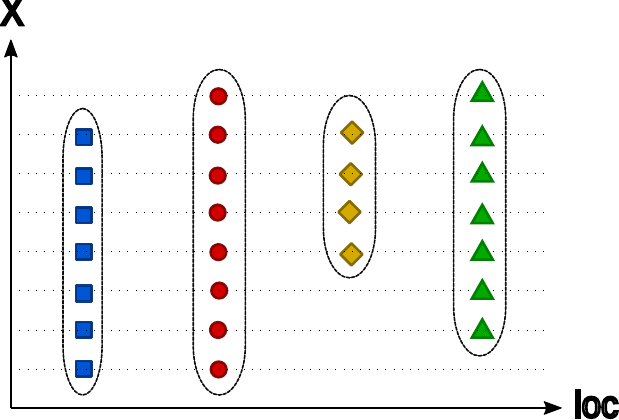}
		\caption{}\label{fig:trace-c}
	\end{subfigure}	
  \caption{From Trace Semantics (a), to State Semantics (b), to Interval Semantics (c).}
  \label{fig:trace}
\end{figure}

The method starts with a concrete semantics providing a precise mathematical expression of program behaviors.
A natural idea is to model a program execution as a discrete sequence of program states.
The trace semantics of a program is then the set of all its possible execution traces.
Figure~\ref{fig:trace-a} depicts, informally, the trace semantics of a program.
A state is composed of the full snapshot of the memory (depicted here simply as the value of variable $X$) and a control location (depicted with different shapes and colors).
A first abstraction consists in collecting, at each location, the set of possible memory states, as depicted in Figure~\ref{fig:trace-b}.
This state semantics forgets about the history of computation (i.e., which state appears before which state), and so, cannot be used to verify temporal properties.
However, it is sufficient (i.e., complete for) state reachability properties, such as the absence of run-time errors.
A further abstraction, the interval semantics, over-approximates the set of values of $X$ at each control location with an interval.
The benefit is that an abstract memory state can be represented in a very compact way, as a pair of a lower and upper bound for $X$ instead of a set of values.
However, we cannot tell whether the values within these bounds are actually reachable or not and, to ensure that we cover all possible program behaviors, we must assume that all these values are possible.
This is represented in Figure~\ref{fig:trace-c}.
With this abstraction, we can no longer prove that $X$ never takes the value 4.

\begin{figure}[ht]
	\centering
	\begin{subfigure}[b]{0.32\textwidth}
		\centering
		\includegraphics[height=3.25cm]{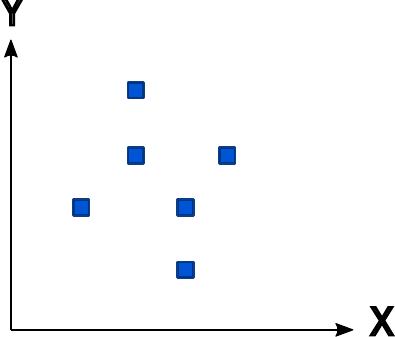}
		\caption{}\label{fig:doms-a}
	\end{subfigure}
	\begin{subfigure}[b]{0.32\textwidth}
		\centering
		\includegraphics[height=3.25cm]{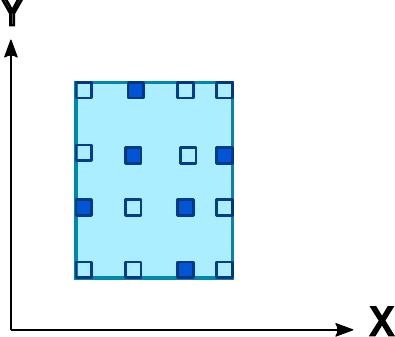}
		\caption{}\label{fig:doms-b}
	\end{subfigure}
	\begin{subfigure}[b]{0.32\textwidth}
		\centering
		\includegraphics[height=3.25cm]{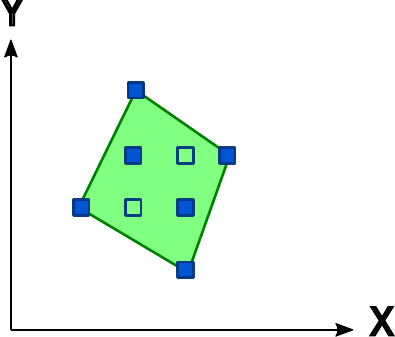}
		\caption{}\label{fig:doms-c}
	\end{subfigure}
  \caption{Concrete Set of Two-Dimensional Points (a), its Interval Abstraction (b), and its Polyhedral Abstraction (c).}
  \label{fig:doms}
\end{figure}

Figure~\ref{fig:doms} gives another example, abstracting a set of points in two dimensions representing variables $X$ and $Y$.
When using the interval abstraction independently on each variable we obtain, in Figure~\ref{fig:doms-b}, an interval with many spurious points (shown as hollow squares in the figure).
The original set-based semantics is undecidable if we assume an infinite range for variable values and, at best, extremely costly to compute when assuming a finite but realistically large range for variables (such as $[-2^{31},2^{31}-1]$).
In the interval semantics, however, we can model assignments using interval arithmetic and tests as constraint programming, which feature efficient algorithms.
Additionally, abstract interpretation introduces fixpoint acceleration operators, called widening, in order to approximate efficiently the limit of iterations involved in the semantics of program loops.
This leads to an efficient static analyzer.

As we perform each computation step within the interval world, each step may induce some loss of precision, which can accumulate.
In practice, the interval analyzer does not infer the tightest possible variable bounds.
It may fail to prove the desired correctness property even when it can be expressed as an interval (such as the absence of overflow): the analysis is not complete and can lead to false alarms.
To achieve the required level of precision, is thus sometimes desirable to use more expressive abstractions.
Figure~\ref{fig:doms-c} presents the classic polyhedral abstraction \cite{PolyhedraDomain}.
On the one hand, it is more precise than intervals (it adds less spurious points) and more expressive (it can infer linear relations between variables, which are particularly useful as loop invariants).
On the other hand, it is much more costly than intervals.
The abstract interpretation literature proposes a large collection of abstract domains, including numeric domains such as intervals and polyhedra, but also domains for pointers, data-structures, etc., which help in finding a good balance between precision and cost.
Abstract interpretation also considers abstractions as first class objects and features operators to combine them.
For instance, the Astrée analyzer \cite{Astree} uses over 40 distinct abstract domains, each one specialized to handle a certain semantic aspect of the proof of absence of run-time errors in embedded critical C code.
This encourages the modular design of static analyzers using reusable parts.

An important benefit of static analysis by abstract interpretation is soundness by construction.
Thus, only the concrete semantics must be trusted, and all subsequent abstractions, including the final computable static analysis, are sound with respect to this concrete semantics.
The concrete semantics is generally a formalization of a standard describing the language (e.g., the C99 standard for Astrée \cite{Astree}), possibly restricted to some implementation choices (such as integer bit-width), that the user and analyzer designer agree upon.
Soundness is generally proved on pen and paper with the help of the abstract interpretation theory, which leaves the possibility of errors in the proofs and bugs in the implementation.
More confidence can be achieved by performing the proofs using a proof assistant, as done for instance in the Verasco analyzer \cite{Verasco}, but this requires a lot of effort for the analysis designer.

\paragraph{Unifying Theory of Formal Methods.}
As a general theory of semantic approximation, abstract interpretation is not only useful to derive new static analyses, but can also model other static analysis methods developed independently, such as data-flow analyses \cite{AIdataflow} or security analyses \cite{AItaint}.
It can also reason about other formal techniques, such as strong typing \cite{AItyping}, constraint programming \cite{AICP}, as well as model checking \cite{AImodelchecking} and SMT solving \cite{AISAT}.
A first benefit of this line of work is theoretical: it uncovers the abstractions that are made, often implicitly, by these techniques.
It helps understanding the expressiveness and the trade-offs made by each method with respect to the most concrete semantic of the program, and provides alternate proofs of soundness and completeness.
In practice, it opens the way to a principled way to combine different formal methods that would seem, at first glance, incompatible.

\looseness=-1
As we will see in the following section, a large part of the formal techniques applied to artificial intelligence employ either SMT solving, constraint solving, or abstract interpretation in a numeric abstract domain.
Hence, abstract interpretation would be a reasonable choice to model, reason about, and combine these techniques.
Additionally, Section~\ref{sec:fm:critical} reported on the industrial use of static analysis by abstract interpretation to verify current critical embedded software.
As machine learning enters critical industries, it would be interesting to combine verification techniques for traditional software and for artificial intelligence using abstract interpretation as a common theory, and develop an analyzer adapted to this new generation of critical software.